\documentclass[longauth]{aa}
\usepackage{graphicx}
\usepackage[colorlinks=true, linkcolor=blue, citecolor=blue, urlcolor=blue]{hyperref}
\usepackage{txfonts}
\usepackage{lscape}
\usepackage{epstopdf}
\usepackage{float}
\usepackage[absolute]{textpos}
\usepackage{amsmath}
\usepackage{mathrsfs}
\usepackage{adjustbox}
\usepackage[T1]{fontenc}
\usepackage{orcidlink}
\usepackage[switch]{lineno}
\usepackage{rotating}
\usepackage{subcaption}

\begin{document}
	
\title{Extreme color--magnitude variability: connection to changing-look AGNs }

\author{
	Litao Zhu\inst{1}\orcidlink{0009-0005-9652-8946}\and
	Zhongxiang Wang\inst{1}\orcidlink{0000-0003-1984-3852}\and
	Alok C. Gupta\inst{2}\orcidlink{0000-0002-9331-4388}\and
	P. U. Devanand\inst{2}\orcidlink{0000-0003-3337-4861}\and
	Ruoheng Yang\inst{1}\and
	Qiangmeng Huang\inst{1}\and
	Man Lang\inst{1}\and
	Jiawen Li\inst{1}
}

\institute{
	Department of Astronomy, School of Physics and Astronomy, Key Laboratory of Astroparticle Physics of Yunnan Province, Yunnan University, Kunming 650091, China\\
	\email{zhulitao@mail.ynu.edu.cn, wangzx20@ynu.edu.cn}
	\and
	Aryabhatta Research Institute of Observational Sciences (ARIES), Manora Peak, Nainital - 263 001, India\\
}

\abstract
{Changing-look active galactic nuclei (CL-AGNs) challenge the unified model of AGNs and offer key insights into the physics of the accretion processes 
of super-massive black holes. While systematic spectroscopic comparisons have 
successfully identified large samples of CL-AGNs, photometric selection based on variability features provides an efficient alternative.}
{We use the color--magnitude (CM) variability pattern method to continue
our identification of the CL transition in AGNs. The resulting CL-AGN sample
	can help our understanding of this phenomenon.}
{The CM variability pattern method utilizes the slope ($k$) of 
the CM variations to identify strong ``bluer-when-brighter'' behavior, while
the variation amplitudes in optical and mid-infrared (MIR) bands are also
considered. The candidates thus selected from
the Type-2 AGNs given in the Sloan Digital Sky Survey (SDSS) catalog are
spectroscopically observed by us
using the 3.6-m Devasthal Optical Telescope (DOT) 
and the 2-m Himalayan Chandra Telescope (HCT). }
{We successfully confirm seven turn-on CL-AGNs among 12 candidates in this work.
Comparing them with both the general AGN populations and the spectroscopically 
identified CL-AGN sample, the CL-AGNs showed larger optical and MIR variations 
and $k$ values.  
The extreme CM variabilities of these sources (with optical magnitude
changes $>$ 0.9) occurred recently. 
For four sources, flare-like brightening
episodes were temporally associated with the turn-on transitions within
3--7 years, suggesting that these flares may trace short-timescale accretion
enhancement, central brightening, and BLR re-illumination.
In addition,
the estimated Eddington ratios for the confirmed CL-AGNs appear to cluster 
around a critical value of $\lambda_{\rm Edd} \sim 0.01$. }
{The extreme CM variability serves as a highly efficient criterion for 
finding CL-AGNs. The properties of the CL-AGNs thus found suggest that
they may represent AGNs at a pivotal state, which likely occur CL transitions
due to enhanced accretion activity, while the cause of the accretion 
activity, determined to have a time scale of several years in this work, 
remains to be further investigated.}


\maketitle
\section{Introduction} \label{sec:intro}

Active galactic nuclei (AGNs) are classified into two major categories based 
on their ultraviolet (UV) and optical spectral features: Type 1 AGNs, which 
exhibit both broad and narrow emission lines, and Type 2 AGNs, which show only 
narrow emission lines. Under the standard unification scheme, this dichotomy 
is governed primarily by an observer's viewing angle relative to an optically 
thick, dusty torus that surrounds the central supermassive black hole (SMBH); 
in Type 2s, their broad-line regions (BLRs) are obscured 
while in Type 1s, their BLRs are visible 
\citep[e.g.,][]{law87, ant93, up95, tad08}. However, this static 
orientation-based paradigm has been challenged by the discovery of the
changing-look (CL) phenomenon:  AGNs can undergo drastic spectral transitions 
between the two types, marked by the appearance/disappearance of 
broad emission lines (BELs) while sometimes involving intermediate subtypes such
as Types 1.2, 1.5, 1.8, and 1.9 (defined by the flux ratios of 
H$\beta$ and H$\alpha$ relative to [O III]; \citealt{ost81, win92}).
The initial instances of the CL transitions were reported decades ago in 
individual sources \citep[e.g.,][]{to76, crp+86, sbw93, ajk+99, eh01, ddc+14, spg+14, lcm+15}. 
The advent of large-area time-domain 
photometric and spectroscopic surveys,
in recent years, have enabled the discovery of a growing 
population of these transition events, transforming 
the studies from accidental discoveries to systematic analyses \citep[e.g.,][]{rac+16, rcr+16, ghc+17, ywf+18, rgc+20, zlw+24, gzf+24, gzg+25}. 

The CL transitions often occur on timescales of a few years, and 
their physical driver is a subject of intense debate.
The standard $\alpha$-disk model predicts that global changes in the accretion 
flow should occur on the viscous timescale, which is typically on the order of 
$10^4$--$10^5$ years for those of SMBHs \citep[e.g.,][]{law18}. While variable 
obscuration by moving dusty clouds offers a possible
solution \citep[e.g.,][]{eli12}, mounting evidence from mid-infrared (MIR) dust 
echoes and continuum reverberation mapping favors the causes due to 
intrinsic variations 
in the accretion rate or state transitions analogous to those in X-ray 
binaries \citep[e.g.,][]{nd18, rae+19,lwl+19,ady+20,grd+25}. Consequently, 
CL-AGNs provide a unique, real-time laboratory to probe the physics of 
accretion instabilities, the activity cycles of active nuclei, and 
the dynamical connection between the central accretion-powered engine and 
the BLR.

\begin{table*}
	\centering
	\caption{Variability metrics for all observed sources.}
	\label{tab:delta-mags}
	\begin{tabular}{lccccccc}
		\hline
		Name & $\Delta zg$ & $\Delta zr$ & $\Delta W1$ & $\Delta W2$ & $k$ & $\rho_{\rm S}$ & $p_{\rm S}$\\
		\hline
		\multicolumn{8}{l}{\textbf{Confirmed CL AGN}}\\
		\hline
		J014638+131109 & $1.249 \pm 0.062$ & $0.864 \pm 0.035$ & $0.359 \pm 0.022$ & $0.480 \pm 0.045$ &$0.46\pm0.02$ & $0.324$ & $4.6\times10^{-10}$\\
		J115953+051330 & $0.963 \pm 0.022$ & $0.714 \pm 0.018$ & $0.697 \pm 0.011$ & $0.927 \pm 0.017$ &$0.27\pm0.02$ & $0.428$ & $5.0\times10^{-7}$\\
		J121228+501412 & $1.151 \pm 0.082$ & $0.785 \pm 0.042$ & $0.667 \pm 0.016$ & $0.930 \pm 0.031$ &$0.20\pm0.03$ & $0.152$ & $6.0\times10^{-4}$\\
		J131101+000310 & $1.042 \pm 0.032$ & $0.769 \pm 0.022$ & $0.629 \pm 0.014$ & $0.849 \pm 0.025$ &$0.55\pm0.02$ & $0.388$ & $2.9\times10^{-8}$\\
		J134330+510204 & $1.094 \pm 0.044$ & $0.694 \pm 0.026$ & $0.647 \pm 0.011$ & $0.784 \pm 0.019$ &$0.46\pm0.01$ & $0.140$ & $1.1\times10^{-4}$\\
		J152502+110744 & $1.723 \pm 0.063$ & $1.075 \pm 0.042$ & $0.755 \pm 0.015$ & $0.643 \pm 0.021$ &$0.641\pm0.007$ & $0.861$ & $1.6\times10^{-10}$\\
		J160626+044802 & $1.297 \pm 0.060$ & $0.971 \pm 0.029$ & $0.467 \pm 0.015$ & $0.677 \pm 0.035$ &$0.45\pm0.02$ & $0.338$ & $3.6\times10^{-10}$\\
		\hline
		\multicolumn{8}{l}{\textbf{Other candidate}}\\
		\hline
		J075846+270515 & $1.042 \pm 0.025$ & $0.804 \pm 0.018$ & $0.616 \pm 0.013$ & $0.993 \pm 0.054$ &$0.250\pm0.007$ & $0.537$ & $3.9\times10^{-10}$\\
		J105344+492955 & $0.945 \pm 0.023$ & $0.648 \pm 0.018$ & $0.201 \pm 0.011$ & $0.618 \pm 0.060$ &$0.22\pm0.01$ & $0.284$ & $9.6\times10^{-10}$\\
		J111137+140118 & $1.219 \pm 0.053$ & $0.747 \pm 0.041$ & $0.610 \pm 0.020$ & $0.644 \pm 0.034$ &$0.48\pm0.02$ & $0.603$ & $1.7\times10^{-10}$\\
		J111614+061736 & $1.029 \pm 0.034$ & $0.819 \pm 0.022$ & $0.761 \pm 0.015$ & $1.099 \pm 0.037$ &$0.37\pm0.02$ & $0.572$ & $5.0\times10^{-10}$\\
		J132558+411500 & $2.656 \pm 0.062$ & $1.982 \pm 0.038$ & $1.824 \pm 0.017$ & $1.885 \pm 0.025$ &$0.185\pm0.003$ & $0.785$ & $1.9\times10^{-10}$\\
		\hline
\end{tabular}
\end{table*}

Systematic searches for CL-AGNs have primarily followed two distinct pathways. 
The first relies on the direct comparison of multi-epoch spectra, from either 
cross-matching data of different 
surveys \citep[e.g.,][]{ywf+18, gzf+24, dzg+25, gzg+25,lkc+26} 
or analyzing repeated observations within a single 
survey \citep[e.g.,][]{gpa+22, zte+24, cjg+25}. While definitive, 
this approach is limited by the scarcity 
of spectroscopic epochs. The second involves pre-selections of candidates 
based on photometric variability characteristics,
which screen for high-amplitude variations in optical 
\citep[e.g.,][]{mrl+16, fgg+19, grs+20, wwg+24, ygw+25,vfb+26} 
or MIR bands \citep[e.g.,][]{swj+20, wzb+23},
followed with spectroscopic observations for confirmation.
Such efforts recently have also expanded to applications of machine learning 
algorithms \citep[e.g.,][]{lmb+22, lsa+23}. 

Among these pre-selection methods, we have developed 
a color--magnitude (CM) pattern method 
that combines variability amplitude with the CM slope 
(parameterized as $k$)\footnote{The CM slope is defined as $zg-zr=k\,zr+c$, where $zg$ and $zr$ are magnitudes in two bands, obtained nearly simultaneously, and $c$ is the intercept.}, 
By explicitly targeting sources that follow a 
steep bluer-when-brighter (BWB) trend in the optical CM space, this method 
can effectively identify high-probability candidates for turn-on 
CL-AGNs \citep{zwd+25}.
The photometric data used to construct the CM diagrams for candidate selections
were mainly from the Zwicky Transient Facility (ZTF; \citealt{bkg+19}),
which started its survey from approximately 2018. Thus, we were
essentially monitoring a large number of
AGNs with the ZTF light-curve data to select those AGNs showing
significant variations (optical changes $>$0.9\,mag) and strong
BWB trends ($k\geq 0.1$; \citealt{zwd+25}). 
The CL transitions we have identified, as a result, likely 
occurred after 2018, i.e., in less than 8 years.


Following this idea, we continued our CL-AGN identification work by applying
the CM variability pattern method. By conducting spectroscopic 
observations in 2024--2025 
and comparing the obtained spectra with those from the Sloan Digital 
Sky Survey (SDSS; \citealt{aaa+09}), seven CL-AGNs were identified from 
12 selected candidates. Among the identified, two (J115953+051330 and
J134330+510204) were reported by \citet{gzg+25} as CL-AGNs but we did not
know at the times of our observations. Furthermore, by adding spectra
for comparison from
the Large Sky Area Multi-Object Fiber Spectroscopic Telescope (LAMOST; 
\citealt{lamost12}) and the Dark Energy Spectroscopic
Instrument (DESI; \citealt{desi16}), we were able to pin down the occurrence
times for four cases of the CL transitions to within 3--7 years, during 
which flaring activity with
extreme CM variability was seen. This identification study thus 
provides a
potential clue for the connection between the extreme variability and the CL 
transition.
The paper is organized as follows: Sec.~\ref{sec:data} describes the candidate
CL-AGN
selection and data reduction; Sec.~\ref{sec:resu} presents the spectroscopic 
confirmation results and related property-analysis results; Sec.~\ref{sec:disc} 
discusses details in this identification work and possible physical 
interpretations to the results.
Throughout this paper, we adopted cosmological parameters from the Planck mission \citep{paa+18}, \emph{$H_0$} = 67.4\,km\,s$^{-1}$ Mpc$^{-1}$ and $\Omega_m$ = 0.315.

\section{Selection of candidate CL-AGNs and Observations} \label{sec:data}

\subsection{Candidate CL-AGN selection}
We adopted the CM variability pattern method proposed 
in \citep{zwd+25} to select CL-AGN 
candidates. This method relies on the CM slope parameter, $k$,
derived from 
the linear fit of $zg - zr$ vs. $zr$, where $zg$ and $zr$ are the
$g$- and $r$-band magnitudes of a source from the ZTF survey and $zg - zr$ was
calculated from the two magnitudes obtained within one day. 
The linear regression was performed as a weighted least-squares fit 
with the Python package \texttt{iminuit}; the uncertainty of each color value, 
estimated from the ZTF photometric errors, was used as the fit weight, and 
the quoted uncertainty of $k$ is the corresponding $1\sigma$ fit error.
A large $k$ value indicates an AGN exhibiting strong BWB variability,
a feature characteristic of Type 1 AGNs.

The parent sample was constructed based on the spectral classifications of AGNs
from the Data Release 16 (DR16; \citealt{dr16}) of the SDSS. In addition to 
the CM diagrams constructed from utilizing the ZTF photometric data, 
MIR light curves from the 
the \textit{Near-Earth Object WISE} (NEOWISE; \citealt{mbg+11,neowise}) data
from \textit{the Wide-field Infrared Survey Explorer} (WISE; \citealt{wem+10})
were also constructed. Specifically the MIR light curves are at
WISE $W1$ (3.4\,$\mu$m) and $W2$ (4.6\,$\mu$m) bands, which presumably trace 
the dust echo responses to optical variability of AGNs.

Focusing on the spectroscopically-identified Type 2 sources in the sample
(for details see \citealt{zwd+25}), we selected sources with
more significant variability as the candidates, which likely have undergone 
CL transitions.  As we have learned from the previous work in \citet{zwd+25}, 
the requirements were the magnitude change in 
$zg$-band to be $\Delta zg > 0.9$\,mag and in W2-band to 
be $\Delta W2 > 0.4$\,mag.
The selection resulted in 12 candidates for our spectroscopic confirmation
observations. In the selection, we did not consider K-corrections;
the redshifts of the sources generally are not high and the selection results 
are not changed by whether or not we add the K-corrections.
The variability properties of these selected sources are summarized in
Table~\ref{tab:delta-mags}. The last two columns of the table provide
the Spearman's rank
correlation coefficient ($\rho_{\rm S}$) and the corresponding two-sided
$p_{\rm S}$ value for each CM relation. The positive $\rho_{\rm S}$ values and
small $p_{\rm S}$ values served as a check for supporting the
reliability of the fitted CM slopes $k$ that were used in our selection.

\begin{figure*}[htbp]
	\centering
	\includegraphics[width=0.49\textwidth]{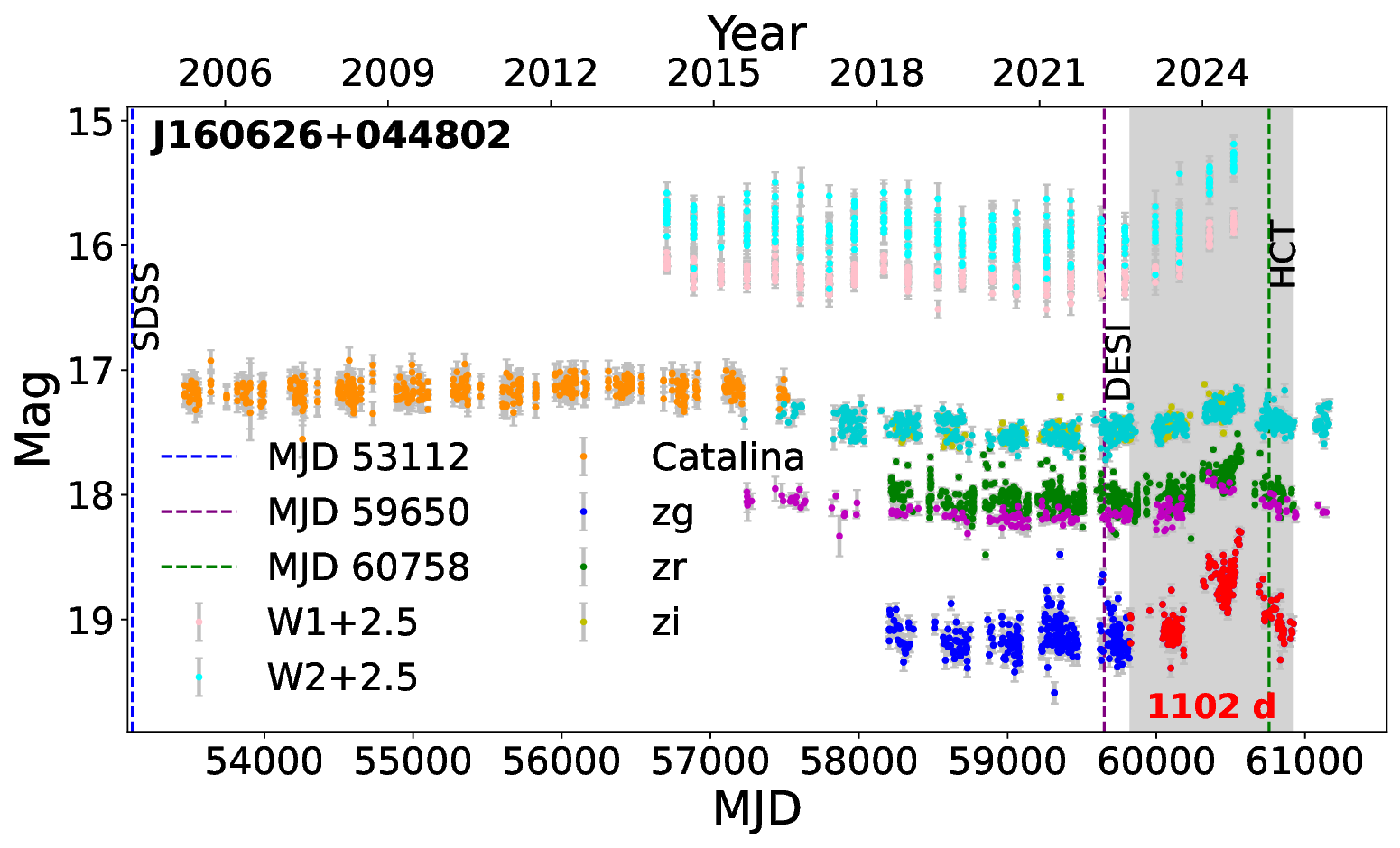}
	\hfill
	\includegraphics[width=0.49\textwidth]{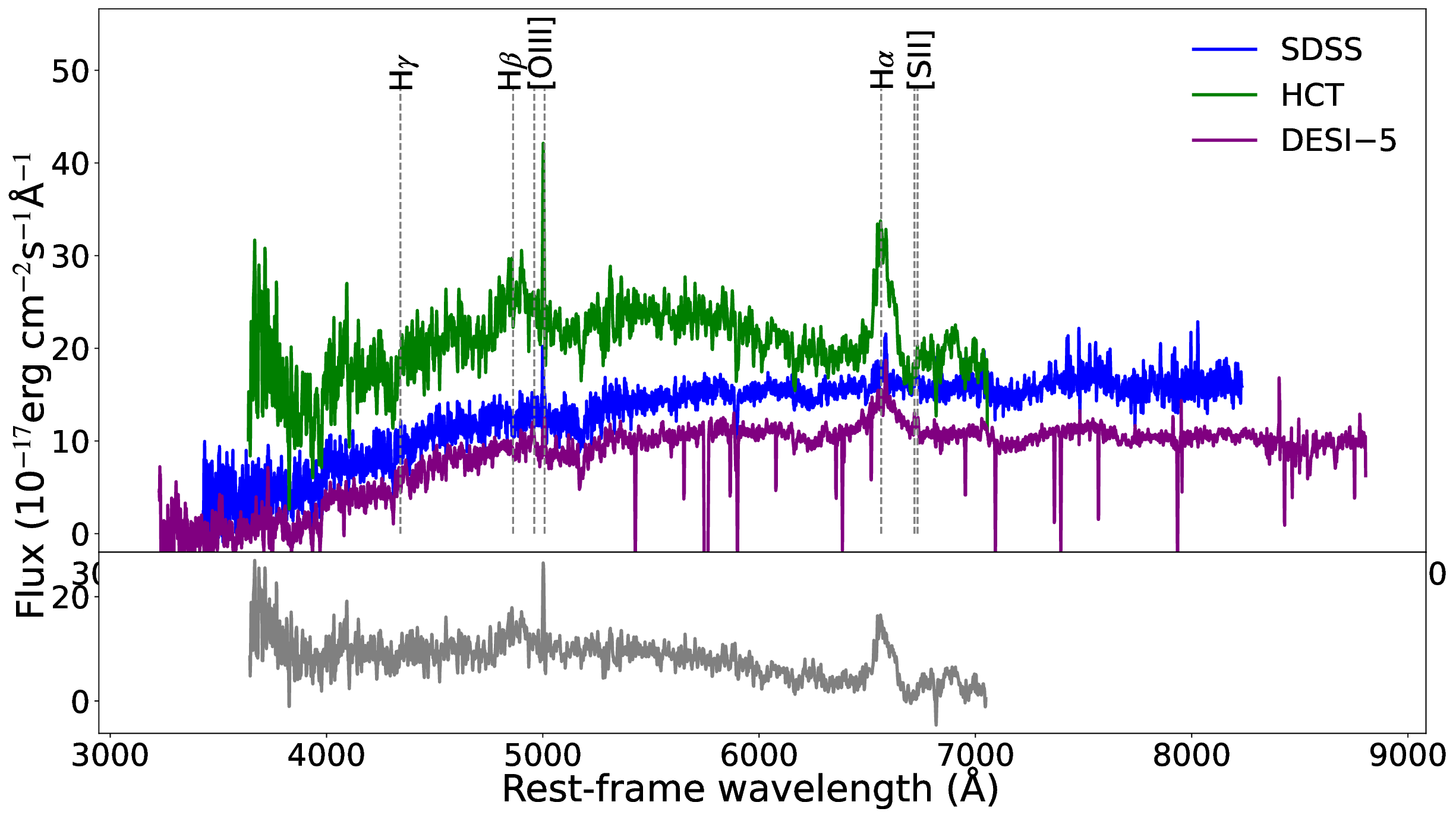}
\caption{Light curves and spectra for J160626+044802. The left panel displays 
	the optical and MIR light curves, where a flare determined from
the $zg$-band data is marked red with its time duration marked by 
the grey region.  
The right top panel shows the spectra obtained from SDSS, DESI, and HCT 
(whose epochs are marked by the vertical dashed lines in the left panel),
and the right bottom panel shows the difference spectrum
	(ours minus the SDSS spectrum; the latter has been convolved with 
	the spectral resolution of ours). }
	\label{fig:J1606}
\end{figure*}

\begin{figure*}[htbp]
	\centering
	\includegraphics[width=0.49\textwidth]{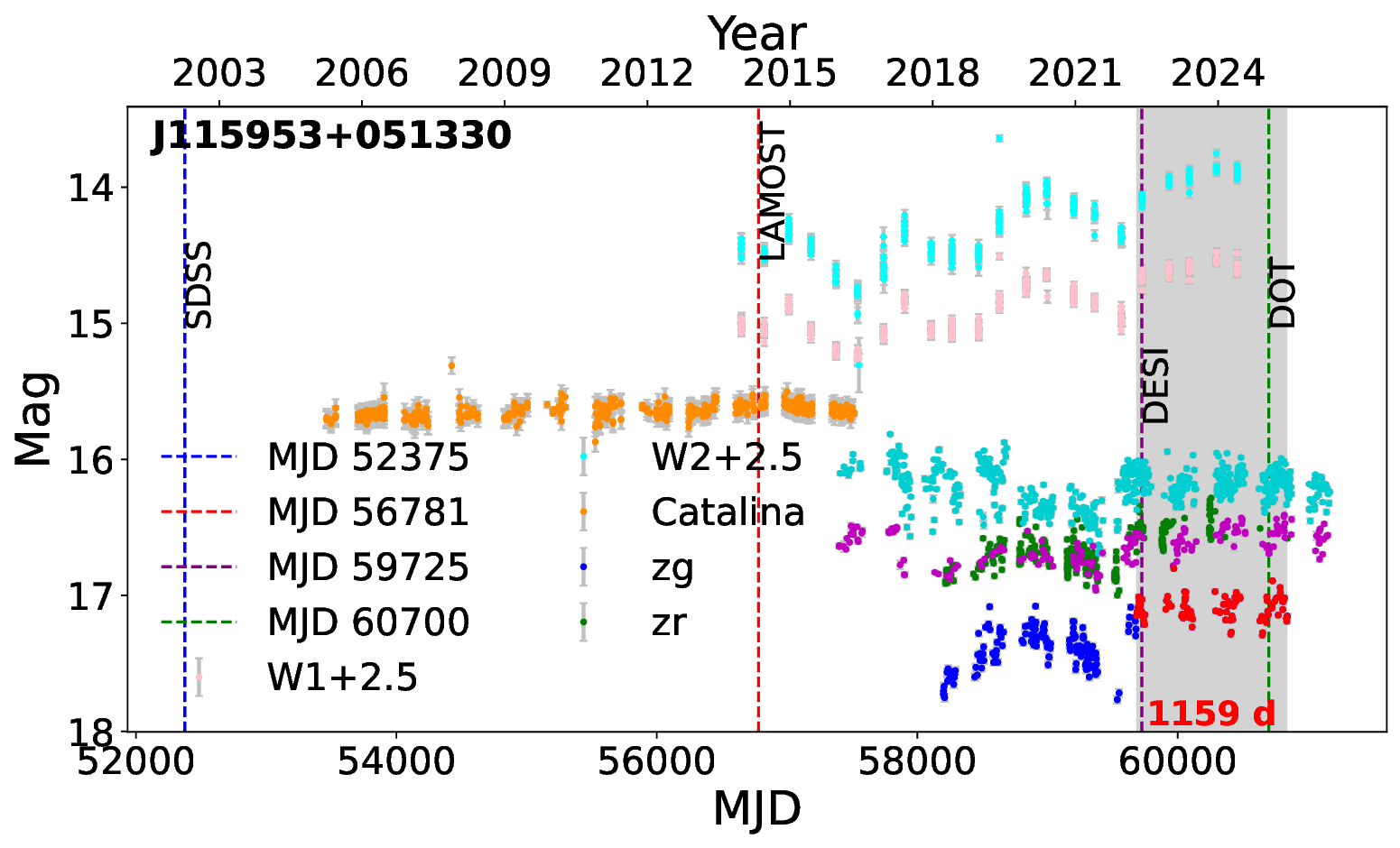}
	\hfill
	\includegraphics[width=0.49\textwidth]{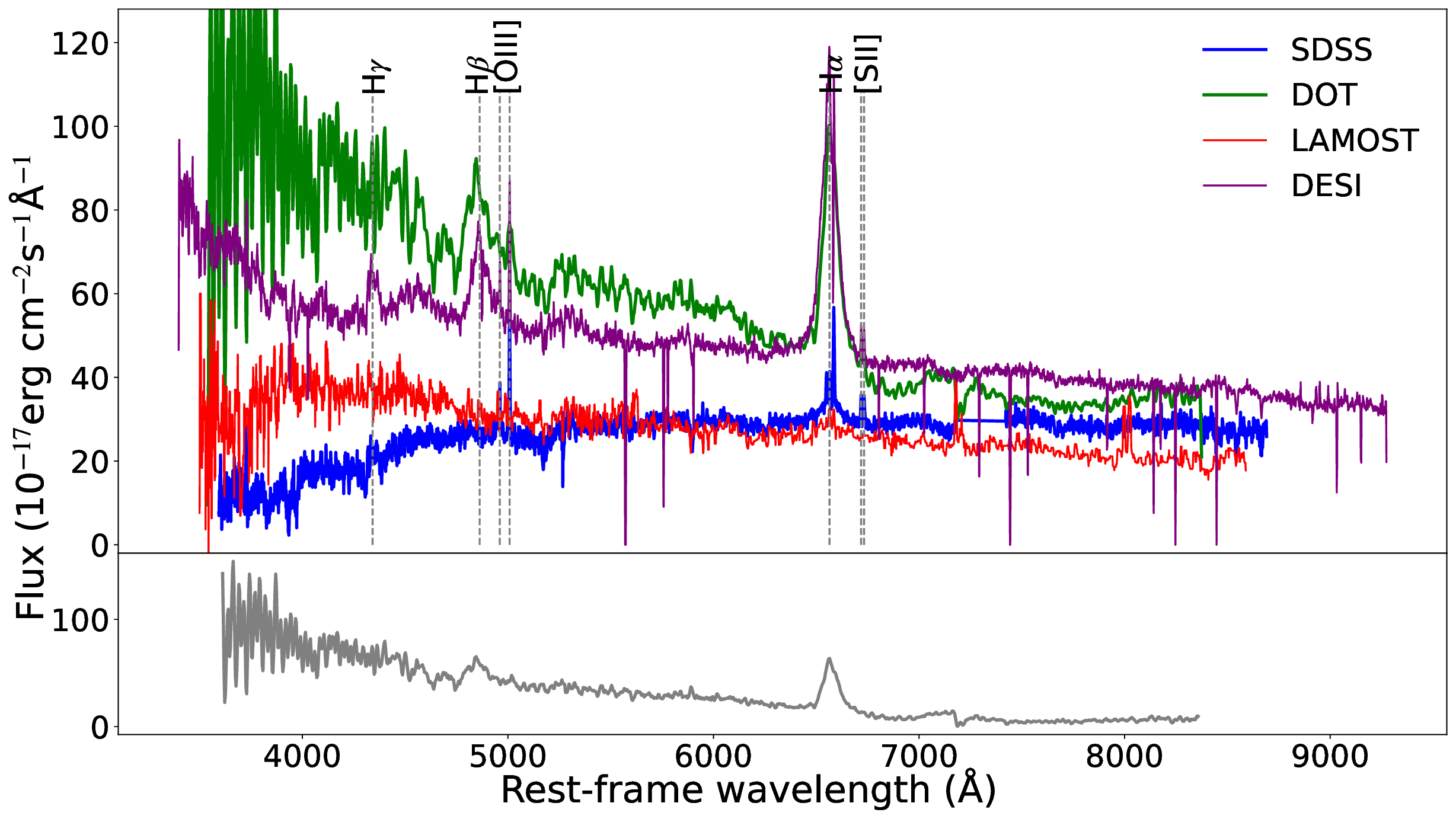}
	\caption{Same as Fig.~\ref{fig:J1606} for J115953+051330, except that 
	the bright spectrum was from the DOT observation and an additional
	LAMOST spectrum was found.}
	\label{fig:J1159}
\end{figure*}

\subsection{Spectroscopic observations and spectral analysis}
\subsubsection{Observations and data reduction}
Spectroscopic observations of the 12 targets were conducted using 
two telescopes: the 3.6-m Devasthal Optical Telescope (DOT; \citealt{okk+19}) 
and the 2-m Himalayan Chandra Telescope (HCT; \citealt{csp02}). The information
for the observations is given in Table~\ref{tab:obs}.

For the observations with DOT, we utilized the ARIES-Devasthal Faint Object 
Spectrograph and Camera (ADFOSC). The instrument is equipped with 
a $4096 \times 4096$ pixel CCD detector. We employed the 132R-600gr/mm grism, 
which provides a spectral dispersion of $\sim 0.10\,\mathrm{nm\,pixel^{-1}}$ 
and covers a wavelength range of 350--700\,nm. A long slit with a width 
of $2.0\arcsec$ was used. Wavelength calibration was performed using spectra
of an Iron-Neon-Argon lamp.

The observations with HCT were carried out using the Hanle Faint Object 
Spectrograph and Camera (HFOSC). To ensure the coverage of key BELs such 
as H$\alpha$ and H$\beta$, we set up the instrument based on the specific 
requirements for obtaining a full spectrum for each target (given its redshift
$z$). We employed Grism 7 (covering $\sim$3800--6840\,\AA) 
paired with a slit of width $0\farcs77$, Grism 8 
(covering $\sim$5800--8350\,\AA) paired with a slit of width $1\farcs15$, 
or a combination of both settings when 
necessary. Similar to the DOT observations, the wavelength calibration for 
the HCT spectra was achieved using the spectra of an Iron-Neon-Argon 
lamp.

The raw spectroscopic data were reduced following the standard procedures by
using the \textsc{IRAF} software package. The reduction steps included 
bias subtraction, flat-fielding, spectral extraction, wavelength calibration, 
and flux calibration. The standard stars observed and used for flux calibration
are listed in Table~\ref{tab:obs}.

\subsubsection{Spectrum data and analysis} 

We downloaded the SDSS spectra, which were mostly taken more than ten years
ago. In addition, we searched the LAMOST and DESI data for our candidates
and found a few spectra that were taken more recently.
The obtained spectra for two confirmed CL-AGNs are shown in 
Fig.~\ref{fig:J1606} \& \ref{fig:J1159} as the examples.

\begin{figure}[htbp]
	\includegraphics[width=0.5\textwidth]{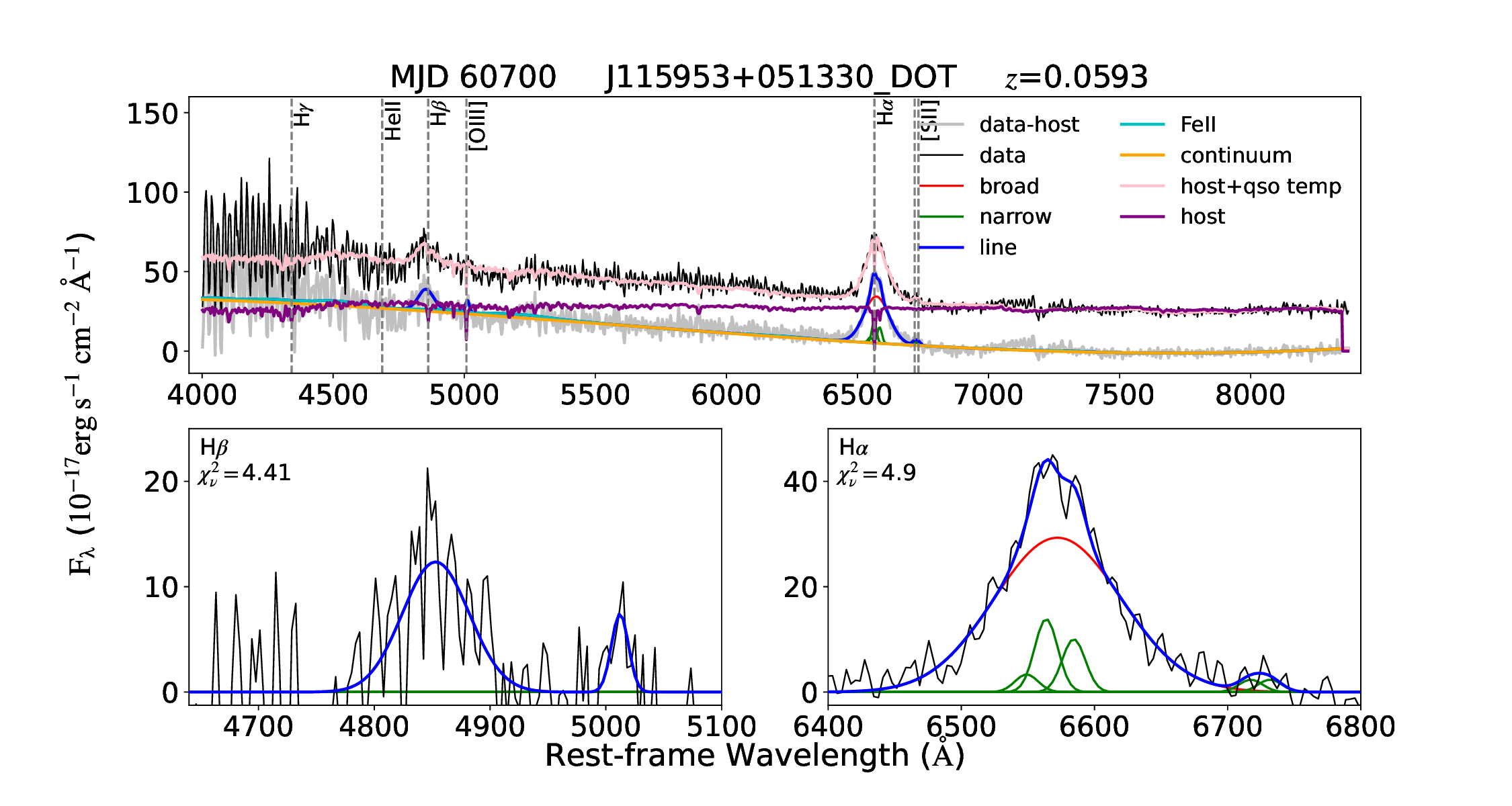}
	\caption{Example of spectral fitting using \texttt{QSOFITMORE} for 
the DOT spectrum of J115953+051330. The spectrum was corrected for redshift 
and Galactic extinction, and the host galaxy contribution was subtracted to 
model the pure quasar emission components and to derive the corresponding 
parameters. The parameters $M_{BH}$ and $\lambda_{Edd}$ for this AGN 
calculated from the fitting parameters are given in Table~\ref{tab:parameter}.}
	\label{fig:J1159_fit}
\end{figure}

To accurately measure the spectral properties of all the spectra, we performed 
spectral fitting using the Python-based package \texttt{QSOFITMORE} 
\citep{qsofitmore}. 
This package, which extends the \texttt{PyQSOFit} 
framework \citep{gsw18},
was employed to model the continuum and decompose the emission lines. We 
obtained precise measurements of the fluxes and full width at half maximum 
(FWHM) for key BELs, specifically H$\alpha$ and H$\beta$. An example of the
spectral fitting is presented in Fig.~\ref{fig:J1159_fit}.

For the sources exhibiting a detectable broad H$\beta$ component, we estimated 
the single-epoch virial black hole (BH) mass ($M_{\mathrm{BH}}$) using 
the empirical scaling relation provided by \citet{vp06}. The estimation 
follows the formula:
\begin{equation}
	\log \left(\frac{M_{\mathrm{BH}}}{M_{\odot}}\right) = \log \left[\left(\frac{\mathrm{FWHM}(\mathrm{H} \beta)}{\mathrm{km\,s}^{-1}}\right)^{2}\left(\frac{L_{5100}}{10^{44}\,\mathrm{erg\,s}^{-1}}\right)^{0.5}\right] + 0.91,
\end{equation}
where $L_{5100}$ represents the monochromatic luminosity at 5100\,\AA, 
and $\mathrm{FWHM}(\mathrm{H} \beta)$ is the line width of the broad H$\beta$ 
component.
Subsequently, we derived the Eddington ratio, defined 
as $\lambda_{\mathrm{Edd}} = L_{\mathrm{bol}}/L_{\mathrm{Edd}}$,
where the Eddington luminosity is 
$L_{\mathrm{Edd}} = 1.3 \times 10^{38} (M_{\mathrm{BH}}/M_{\odot})\,\mathrm{erg\,s}^{-1}$.
The bolometric luminosity was estimated as 
$L_{\mathrm{bol}} = 9.26 \times L_{5100}$ by adopting the 
scaling from \citep{rls+06} for quasars with $z < 0.8$.

\subsection{Variability analysis}

Our selection was primarily based on variability features.
To display the long-term variability behavior (e.g., Fig.~\ref{fig:J1606} \&
Fig.~\ref{fig:J1159}), we 
included archival photometric data from two additional surveys: 
the $V$-band data 
from the Catalina Real-time Transient Survey (CRTS; \citealt{ddm+09})
and the cyan ($ac$; 420--650\,nm) and orange ($ao$; 560--820\,nm) band data 
from the Asteroid Terrestrial-impact Last Alert System 
(ATLAS; \citealt{tdh+18}). The constructed light curves  
of the five confirmed CL-AGNs and five other candidates are provided in 
the Appendix (Fig.~\ref{fig:J0146}--\ref{fig:J1525} and 
Fig.~\ref{fig:J0758}--\ref{fig:J1325}, respectively).

The long-term light curves display that the CL-AGNs have undergone 
significant flux variations in both optical and MIR bands. To quantify these 
variations, we calculated the maximum amplitude changes: $\Delta zg$, 
$\Delta zr$, $\Delta W1$, and $\Delta W2$, which are summarised in 
Table~\ref{tab:delta-mags}. For the MIR data, the magnitude changes 
were derived using binned WISE light curves, where data points from one
epoch were averaged into one bin. As shown in the table, the CL-AGNs 
consistently exhibited substantial amplitudes in $\Delta zg$, $\Delta zr$, 
as well as in the MIR bands.
The basic properties and variability metrics are summarised in 
Table~\ref{tab:delta-mags}.

Careful examining the light curves of the candidates, we approximately 
identified three types of variations:  a major
flare as in those of 
J115953+051330 (Fig.~\ref{fig:J1159}), J152502+110744 (Fig.~\ref{fig:J1525}),
and J160626+044802 (Fig.~\ref{fig:J1606}), one or two intermediate flares as 
in those of J132558+411500 (Fig.~\ref{fig:J1325}) and 
J134330+510204 (Fig.~\ref{fig:J1343}), and flickers as in those
of J014638+131109 (Fig.~\ref{fig:J0146}) and 
J121228+501412 (Fig.~\ref{fig:J1212}).
In order to understand the role of these variations (in particular
the flares) in forming the CM patterns,
we applied the Bayesian Blocks (BB) algorithm \citep{snj+13}, 
implemented through the Python package Astropy\footnote{https://docs.astropy.org/en/stable/api/astropy.stats.bayesian\_blocks.html},
to $zg$-band 
light curves. The median magnitude of an entire light curve was defined as 
the quiescent baseline. A continuous sequence of blocks significantly brighter 
than this baseline determined a time interval of a flare. From this analysis
we obtained the time interval of a flare for each candidate. The determination
relied on the baseline level and the most deviating data point above
the baseline, which could result in extreme cases such as
a short, 169\,d flare at the end of the light curve of 
J121228+501412 (Fig.~\ref{fig:J1212}). In any case, since we were interested 
in checking
how flares or flickers contributed to the CM patterns, we did not adjust 
the determination for possibly more reasonable results.

To construct CM diagrams, we binned the ZTF $zg$- and $zr$-band data points
over 3-day interval, which helped avoid sometimes one data point of a band
versus multiple data points of another band. The flares determined above from
the BB algorithm were marked in the CM diagrams (Fig~\ref{fig:clcm} \& \ref{fig:nclcm}).

\section{Results}  \label{sec:resu}

\begin{figure*}
	\centering
	\includegraphics[width=0.7\linewidth]{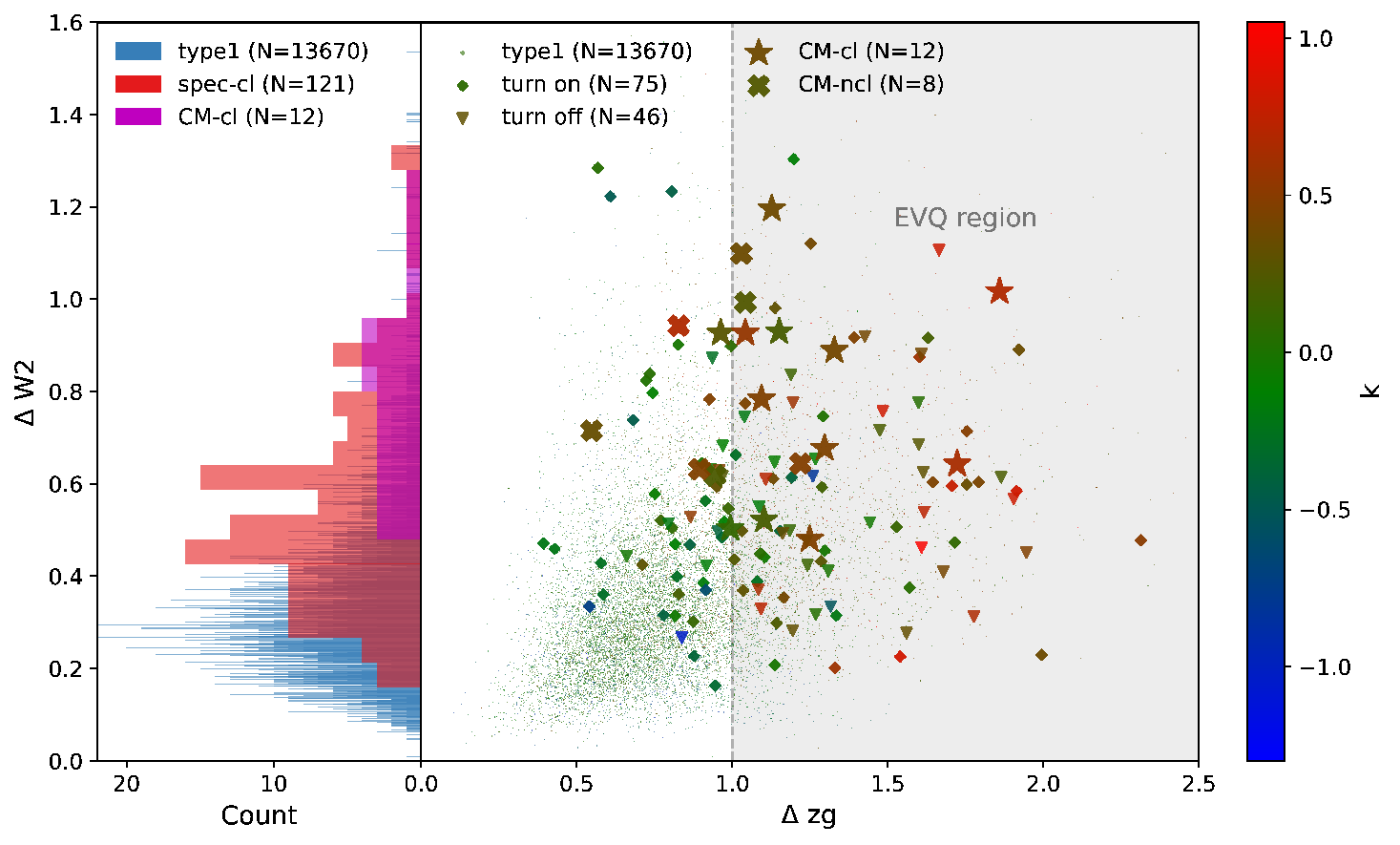}
	\includegraphics[width=0.7\linewidth]{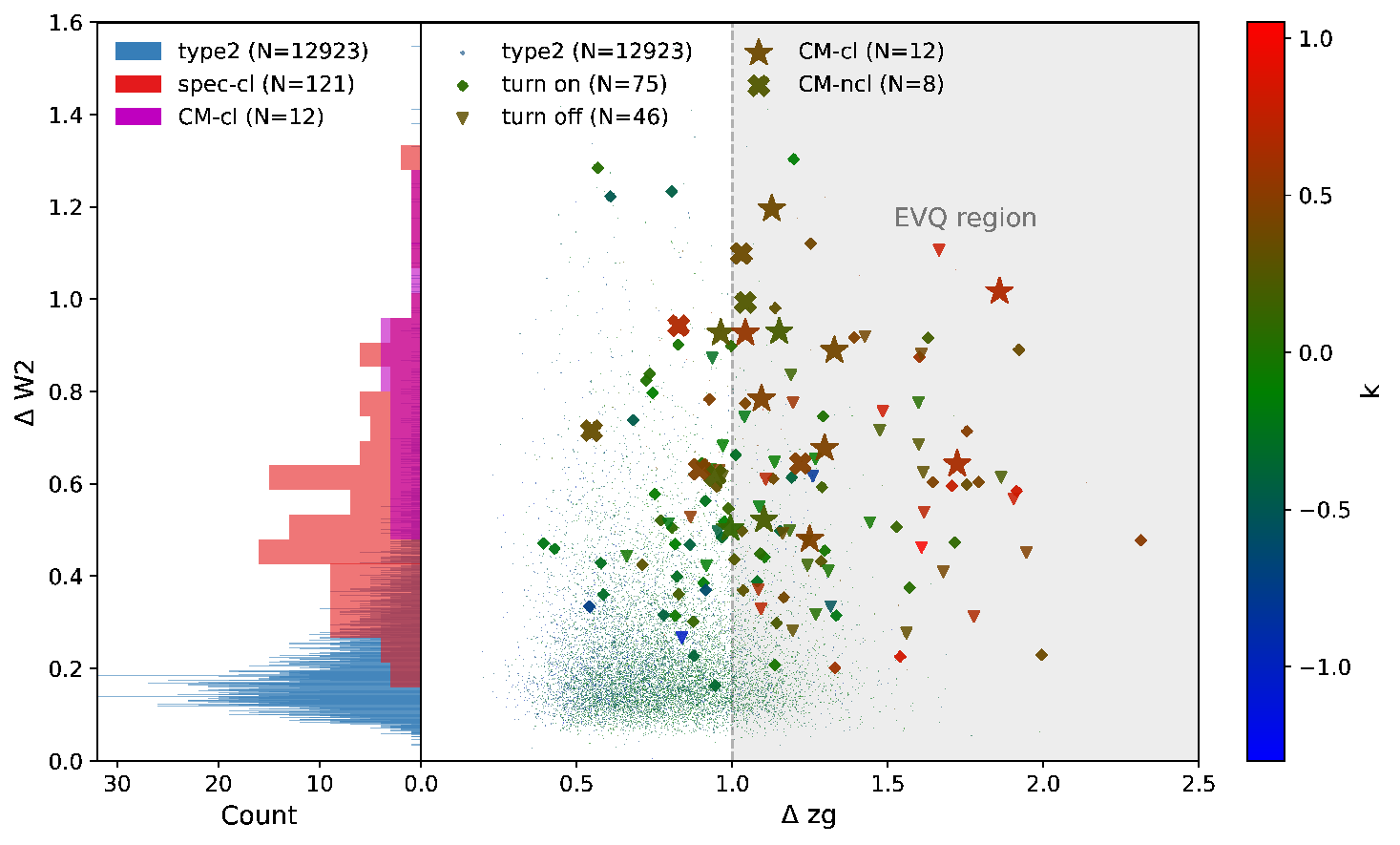}
	\caption{Distributions of variability amplitudes ($\Delta W2$ versus 
$\Delta zg$) and the color--magnitude slope ($k$). The upper and lower panels 
display the general SDSS Type 1 and Type 2 AGN populations as the background, 
respectively. The color bar on the right of each panel represents 
the $k$ value. The histograms on the left side of each panel illustrate 
the distributions of $\Delta W2$. In the plots, ``CM-cl'' and ``CM-ncl'' 
denote the 12 confirmed CL-AGNs and the 8 non-confirmed CL candidates, 
respectively, selected via the CM variability pattern method 
	(including \citealt{zwd+25}
and this work). The samples labeled ``spec-cl'', ``turn on'', and ``turn off'' 
are derived from the currently largest spectroscopically-selected CL-AGN 
	catalog from \citet{gzg+25}.}
	\label{fig:statistic}
\end{figure*}

\begin{table*}
	\centering
	\caption{Properties of seven CL sources.}
	\label{tab:parameter}
	\begin{tabular}{lccccccccc}
		\hline
		Source &  $z$ & Log($M_{BH}$/$M_{\odot}$) & Log($\lambda_{Edd}$) & Transition  \\
		\hline
		J014638+131109  & 0.1601 & $8.17 \pm 0.11$ & $-1.80 \pm 0.13$ & H$\beta$ on \\
		J115953+051330   & 0.0593 & $7.78 \pm 0.17$ & $-1.86 \pm 0.19$ & H$\beta$ on \\
		J121228+501412  & 0.1700 & $8.07 \pm 0.17$ & $-1.89 \pm 0.19$ & H$\beta$ on \\
		J131101+000310  & 0.0962 & $8.24 \pm 0.19$ & $-1.72 \pm 0.23$ & H$\alpha$, H$\beta$ on \\
		J134330+510204  & 0.1316 & - & - & H$\alpha$ on \\
		J152502+110744  & 0.3326 & $8.42 \pm 0.13$ & $-1.54 \pm 0.15$ & H$\beta$ on \\
		J160626+044802  & 0.1158 & $8.29 \pm 0.25$ & $-1.68 \pm 0.29$ & H$\alpha$, H$\beta$ on \\
		\hline
	\end{tabular}
\end{table*}
Our spectroscopic observations of 12 targets resulted in the successful 
identification of 7 turn-on CL-AGNs. 
Representative examples of the identified CL-AGNs are presented in 
Fig.~\ref{fig:J1606} (observed with HCT) and Fig.~\ref{fig:J1159} 
(observed with DOT).

Brief descriptions of the individual CL-AGN sources are presented below:


\noindent\textbf{J014638+131109:} The early CRTS data for this source 
were relatively stable. A rising trend appeared in the MIR data starting 
from 2016. There was a flux-variation dip appearing clearly in $zg$ band
after 2021, followed with a recovering upward trend. A time period of 
$\sim$680\,d in the first part of the $zg$ light curve was determined
to be a flare.  Our HCT spectrum was obtained during a 
relatively bright phase. Compared to the extremely weak broad H$\alpha$ 
observed in the earlier SDSS epoch, the HCT spectrum revealed prominent 
broad H$\alpha$ and H$\beta$ lines (Fig~\ref{fig:J0146}).

\noindent\textbf{J115953+051330:} The CRTS data for this source did not show 
significant variability. The more recent ATLAS and ZTF data indicated 
fluctuations but the $zg$-band data suggested a long flare before $\sim$2022 and
a stable brightening state after; the latter was identified as a flare
with the BB algorithm. The MIR response was aligned with 
the $z_g$-band variations, likely as the addition to an overall brightening 
trend. The LAMOST spectrum taken before 2015 indicates a Type-2 AGN, and
the DESI (as already reported in \citealt{gzg+25}) and our DOT spectra, 
both obtained during the identified flare, revealed
significantly enhanced broad H$\alpha$, a fully emerged broad H$\beta$, and 
a clear continuum component (Fig~\ref{fig:J1159}). 


\noindent\textbf{J121228+501412:} The MIR light curves of this source 
showed a continuous dimming trend during $\sim$2015--2017.
Starting from $\sim$2019, there was a slow brightening trend particularly 
visible in $zg$ band. Before the brightening, the LAMOST spectrum 
	revealed the clear absence of broad H$\beta$ and H$\gamma$ components,
	which emerged in our HCT spectrum obtained during the identified
	short flare (Fig~\ref{fig:J1212}).
	In addition, comparing the HCT spectrum
	to the much early SDSS spectrum, 
the H$\alpha$ emission turned to be significantly strengthened. The time 
	duration between the LAMOST and HCT observations is $\sim$6.5\,years.

\noindent\textbf{J131101+000310:} The variability of this source was mild in 
the early years until $\sim$2020, when the light curves, particularly 
at $zg$-band, showed sharp intermediate flares, accompanied with
	MIR flux increases. The LAMOST spectrum taken at $\sim$2018
	and our DOT spectrum taken after the identified optical flares
	established a CL transition between the epochs of the two spectra
	(having a time duration of $\sim$7.0 years);
the latter displayed a significantly enhanced broad H$\alpha$ and an 
emerged broad H$\beta$ (Fig~\ref{fig:J1311}). The transition was
likely in association with the flaring activity between the epochs of
	the two spectra.

\noindent\textbf{J134330+510204:} Similar to J131101+000310, this source was 
relatively quiet in early years, but with a noticeable flare occurring
	in $\sim$2023. The DESI spectrum (already reported in \citealt{gzg+25})
	taken right before the identified flare showed emerged broad H$\alpha$
	emission, and the DOT spectrum taken after
	the flare indicated a stronger broad H$\alpha$ component 
	(Fig~\ref{fig:J1343}).

\noindent\textbf{J152502+110744:} The variability of this source was not 
intense in the early years. However in late 2020, the optical flux increased 
significantly, appearing as a major flare with $zg$ magnitude up by 
	$\sim -$1.8 mag. The MIR light curves followed the optical ones. 
	Our HCT spectrum, taken during 
the brightening stage, exhibited a significantly enhanced continuum and 
broad H$\alpha$ component, as well as a potentially very broad H$\beta$ 
component (Fig~\ref{fig:J1525}). The DESI spectrum taken at the onset of
of the flare, which did not show these features, limited the CL transition
within $\sim$4.0 years.

\noindent\textbf{J160626+044802:} This source remained relatively stable in 
both optical and MIR bands until the appearance of a major flare from
$\sim$2022. The early SDSS 
spectrum was dominated by the galaxy component without obvious BELs and
the DESI spectrum taken right before the flare was similar but showing the
appearance of a broad H$\alpha$ component. In our 
HCT spectrum taken after the peak of the flare, distinct broad H$\alpha$ 
and H$\beta$ 
emission lines appeared, accompanied by an enhanced continuum component
(Fig~\ref{fig:J1606}). Similar to J152502+110744, the transition should have
been associated with the flare, occurring within the $\sim$3.0 years between 
the DESI and HCT observations.


To place our results in a broader context, we combined the 7 CL-AGNs 
identified in this work with the 5 CL-AGNs previously identified using 
the same method \citep{zwd+25} and plotted their $\Delta zg$ and $\Delta W2$ 
in Fig.~\ref{fig:statistic}, with their $k$ values (derived from the CM 
variations) marked with a color scale of from blue to red
(corresponding $k$ from $\sim -1$ to $\sim$+1).
For comparison, we also plotted the distribution of Type 1 AGNs 
(left panel of Fig.~\ref{fig:statistic}) and Type 2 AGNs 
(right panel of Fig.~\ref{fig:statistic}) from the SDSS DR16 catalog. 
Furthermore, we included a sample of 121 CL-AGNs from \citet{gzg+25} that 
have available ZTF and WISE light curve data.
As illustrated in Fig.~\ref{fig:statistic}, the Type 1 and Type 2 AGNs are 
clustered in the region characterized by relatively small values 
of $\Delta zg$, $\Delta W2$, and $k$. 
The majority of Type 2 AGNs exhibit $k$ values less than zero, whereas Type 1 
AGNs are distributed around $k \approx 0$ (see also \citealt{zwd+25}). 
For Type 2 AGNs, we would like to point out that some of
their apparent negative $k$ values do not necessarily imply a 
physical ``redder-when-brighter'' behavior; rather, they primarily result 
from unreliable linear fitting caused by the lack of significant intrinsic 
variability in these objects.
The CL-AGNs identified through spectrum comparisons, i.e., 
the sample in \citealt{gzg+25}, occupy a region with larger variability 
amplitudes than those of Type 1/2 AGNs, but notably not as large as those of
the CL-AGNs found from our CM variability method. 

\begin{figure}[htbp]
	\centering
	\includegraphics[width=0.45\textwidth]{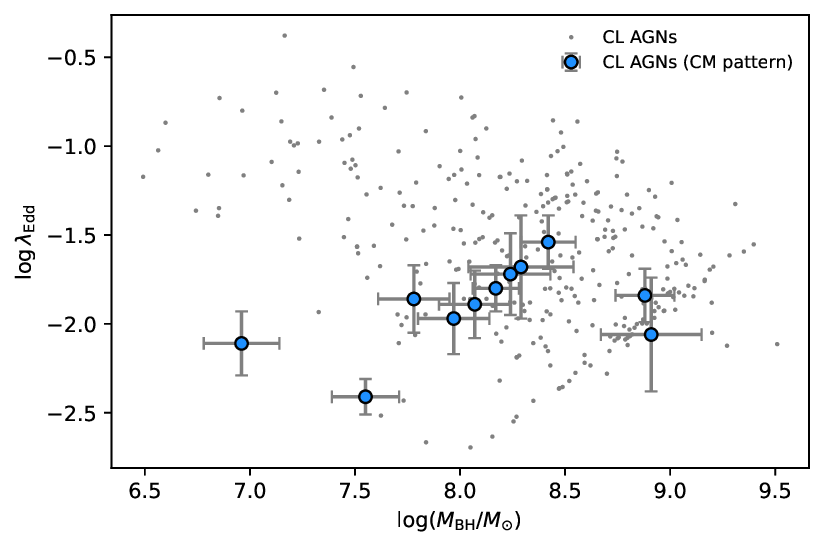}
	\caption{Distribution of black hole masses ($M_{\mathrm{BH}}$) versus Eddington ratios ($\lambda_{\mathrm{Edd}}$) for the CL-AGNs identified via 
the CM variability method. The data points comprise six confirmed 
sources from this work and the four sources previously reported in
\citet{zwd+25}. The background data points are those CL-AGNs reported by
	\citet{gzg+25}.}
	\label{fig:mbh_edd}
\end{figure}

The estimated black hole masses ($M_{\mathrm{BH}}$) and Eddington ratios
($\lambda_{\mathrm{Edd}}$) of six confirmed CL-AGNs (Table~\ref{tab:parameter}),
as well as those of four CL-AGNs reported in \citet{zwd+25}, are shown
in Fig.~\ref{fig:mbh_edd}. For comparison, the values of the CL-AGN
sample reported by \citet{gzg+25} are also plotted in the figure.
This currently largest CL-AGN sample was built by comparing spectrum data from 
the SDSS and DESI.
The ten CL-AGNs identified from the CM variability pattern method 
are in ranges of $\log (M_{\mathrm{BH}}/M_{\odot}) \sim 6.9$ to $9.0$
and $\log \lambda_{\mathrm{Edd}} \sim -2.4$ to $-1.4$.
Thus, these ten AGNs generally have $\lambda_{\mathrm{Edd}}$ at the low end.
This feature of low $\lambda_{\mathrm{Edd}}$ has also been noted, for
example, in extreme variability quasars (EVQs; \citep{rsm+18}) and in 
CL quasars \citep{mac+19}.

\section{Discussion} \label{sec:disc}

Following our CM variability method established in \citet{zwd+25}, we carried 
out further studies and observed another 12 sources in 2024--2025. Seven of
them were confirmed as CL-AGNs (while two of them were already reported
in \citealt{gzg+25}).
A detailed examination of the unconfirmed candidates reveals that several
factors may hinder the successful identification of a CL transition.
First, most of the candidates were selected because they recently exhibited
variations of a flare-like event
(e.g., Fig.~\ref{fig:J1606} \& \ref{fig:J1525}) and
the BWB pattern was dominantly established by this event
(Fig.~\ref{fig:clcm}),
and thus the time lag between the photometric flare and the spectroscopic
follow-up could be critical. For instance, J111614+061736
(Fig.~\ref{fig:J1116}) was observed $\sim$4 years after its light-curve
peak, potentially causing the miss of the broad-line emergence window.
This possibility might also occur in the unconfirmed cases of J1246$-$0156,
J1252+0717, and J1423+2454 reported in \citet{zwd+25}.
Second, data quality can affect the results. The DOT spectra suffer large
uncertainties in the blue part, which is particularly severe for
the identification of broad lines
in J105344+492955 (Fig.~\ref{fig:J1053}).
Similarly, the DOT spectrum of J111137+140118 (Fig.~\ref{fig:J1111}) exhibited 
a significant strengthening of broad H$\alpha$, yet it lacked the unambiguous
emergence of broad H$\beta$; thus, we conservatively excluded it from our
turn-on sample.
Third, careful examination of J132558+411500's light curves and spectra
(Fig.~\ref{fig:J1325}) made us realize that it is likely a blazar
as it exhibited large daily optical variations (over 1\,mag),
nearly simultaneous fast flux variations in MIR bands, and flat continuum plus 
weak lines.
Indeed, it is identified as a blazar of BL Lac type 
(4FGL~J1326.2+4115 or IVS B1323+415) in 
the Fermi Large Area Telescope (LAT) catalog \citep{4fgl-dr4}. A lesson is
learnt that
our selection of candidates requires further checking of multi-band information
in order to exclude sources such as this blazar.

While this work proves again that the CM variability pattern method can 
effectively
find turn-on CL-AGNs, the candidates we selected approximately belong to 
the EVQ type, i.e.
AGNs showing $\geq 1$\,mag changes \citep{rsm+18, rwc+22}. In the 
selection of this work, we also considered the MIR variations
($\Delta W2 >0.4$) and the clear BWB pattern ($k\geq 0.1$).
Among the 12923 SDSS Type-2 AGNs, there were 42 satisfying these requirements.
Because we have access to 2--3-meter telescopes for spectroscopic observations,
the targets selected 
were relatively bright, with averaged $zg$ magnitudes smaller than 19.5.
This brightness requirement excludes seven faint sources.
Among the remaining 35 Type-2 AGNs, we observed 17, of which 12 were identified
as CL-AGNs (including five from \citealt{zwd+25}). Therefore, when a
previously identified Type-2 AGN shows large BWB flux variations,
it is highly likely that it would have undergone a CL transition. 
The cause of the large variations in several of our cases was several-year long
flare-like events,
which may provide a clue to understanding the CL process. 
One possible scenario is that if there is a denser part of the inflow reaching
the inner disk, the accretion rate rises in a short time, producing a 
brighter optical/UV
continuum and more ionizing photons \citep{da11,db19,scb+20}. The stronger
ionizing continuum illuminates more BLR gas, and so the effective BLR size
increases, as expected from the BLR radius--luminosity relation and BLR
breathing behavior \citep{bdg+13,gk14}. Since broad H$\alpha$ is usually
stronger and may arise from a slightly larger, lower-ionization region than
broad H$\beta$, an early or weak turn-on phase may show broad H$\alpha$ while
broad H$\beta$ is still weak or absent (similar to Seyfert Type 1.8/1.9;
\citealt{ost81,win92}). This early turn-on of broad H$\alpha$ was possibly
seen in J131101+000310 (Fig.~\ref{fig:J1311}) and 
J134330+510204 (Fig.~\ref{fig:J1343}).
In this sense, the method we used would
preferentially select high-variability, recent turn-on CL events, in
which the flare-like events serve as an observational sign for 
the accretion enhancement in the central engine.

The relatively low Eddington ratios of these CL-AGNs
(Fig.~\ref{fig:mbh_edd})
may suggest that their previous ``Type 2'' classification was
intrinsic, consistent with the concept of ``naked'' Type-2
AGNs \citep{ygm+23}. In such AGNs, BELs are absent, yet no obscuration is
detected in X-ray observations \citep{lyp+14, jrt+25}. These CL-AGNs
showed the MIR variabilities that closely followed the optical variations,
indicating the absence of obscuration along the line of sight.
The absence of BELs in them was thus likely due to a low accretion rate
insufficient to sustain a visible BLR.
Comparing them to the larger
CL-AGN sample from \citet{cjg+25} whose $\log \lambda_{\mathrm{Edd}}$ span
from $\sim -3.2$ to $-0.8$ (see also \citealt{ps24, zte+24, ygw+25} for
similarly broad distributions), they show a
narrow distribution clustered around the critical value
of $\log \lambda_{\mathrm{Edd}} \approx -2$, which may imply that they
represent a transitional group between
the low-accretion Type-2 phase and the high-accretion Type-1 phase.
Since $\log\lambda_{\mathrm{Edd}} \sim -2$ is generally considered to
represent the theoretical
boundary between the Advection-Dominated Accretion Flow (ADAF) and
the Standard Shakura-Sunyaev Disk (SSD; \citealt{nd18}), this clustering 
suggests
that these sources were likely not in a stable accretion mode, but 
rather in a pivotal state between the ADAF and SSD modes.


As the CL-AGN population continues to expand, a variety of
mechanisms have been proposed.  For example, there are scenarios of
varying obscuration caused by disk winds or
outflows \citep[e.g.,][]{eli12,gh18}, the presence of SMBH
binaries that perturb the accretion flow \citep{wb20}, the propagation of
cooling fronts within the accretion disk \citep{rfg+18}, the inefficient
cooling of hot outflows from ADAFs preventing BLR formation \citep{cao10},
and magnetic or radiation pressure instabilities triggering rapid luminosity
changes \citep[e.g.,][]{db19,scb+20}. It is likely because of the finding
method we used, the CL-AGNs in our sample appear to fit in the accretion-mode
transition scenario. However, even within this framework,
different triggers may be at play. For instance, tidal disruption events (TDEs)
in AGNs can fuel a sudden, dramatic brightening \citep[e.g.,][]{tam+19,bnb+17}. 
Alternatively, \citet{klw+25} proposed that
Lense-Thirring (LT) precession in a misaligned accretion disk can lead
to ``disk tearing,'' causing rapid infall of material and triggering a CL
transition. Crucially, this mechanism is sensitive to the accretion rate. 
At the critical rate of $\lambda_{\rm Edd} \sim 0.01$,
the disk is susceptible to tearing, potentially turning the BLR on or off 
completely. 
Finally, drawing an analogy to stellar-mass BH systems offers a unifying 
perspective. Such studies \citep[e.g.,][]{rae+19,lwl+19,ady+20,grd+25} have 
linked the CL-AGN phenomenon to the high-soft/low-hard state transitions 
observed in X-ray binaries. This scale-invariant view suggests that 
the physics of accretion flow changes---driven by the interplay among the 
components, disk, corona, and jet---may be fundamental across all mass scales, 
providing a robust framework for understanding the extreme-variability--related
CL transitions.

\section*{Data availability}
The derived data underlying this article are available from the corresponding author on reasonable request.

\begin{acknowledgements} 
This work was based on observations obtained with the Samuel Oschin Telescope 48-inch and the 60-inch Telescope at the Palomar Observatory as part of the Zwicky Transient Facility project. ZTF is supported by the National Science Foundation under Grant No. AST-2034437 and a collaboration including Caltech, IPAC, the Weizmann Institute for Science, the Oskar Klein Center at Stockholm University, the University of Maryland, Deutsches Elektronen-Synchrotron and Humboldt University, the TANGO Consortium of Taiwan, the University of Wisconsin at Milwaukee, Trinity College Dublin, Lawrence Livermore National Laboratories, and IN2P3, France. Operations are conducted by COO, IPAC, and UW.\\

This study makes use of data obtained from the 2-m Himalayan Chandra Telescope (HCT). We thank the staff at IAO, Hanle, and CREST, Hosakote, operated by the Indian Institute of Astrophysics, Bengaluru (India), for their support in facilitating these observations.\\
	
This work is based on observations obtained at the 3.6m Devasthal Optical Telescope (DOT), which is a National Facility run and managed by Aryabhatta Research Institute of Observational Sciences (ARIES), an autonomous Institute under the Department of Science and Technology, Government of India.\\
	
Funding for the Sloan Digital Sky Survey V has been provided by the Alfred P. Sloan Foundation, the Heising-Simons Foundation, the National Science Foundation, and the Participating Institutions. SDSS acknowledges support and resources from the Center for High-Performance Computing at the University of Utah. SDSS telescopes are located at Apache Point Observatory, funded by the Astrophysical Research Consortium and operated by New Mexico State University, and at Las Campanas Observatory, operated by the Carnegie Institution for Science. The SDSS web site is \url{www.sdss.org}.\\

SDSS is managed by the Astrophysical Research Consortium for the Participating Institutions of the SDSS Collaboration, including the Carnegie Institution for Science, Chilean National Time Allocation Committee (CNTAC) ratified researchers, Caltech, the Gotham Participation Group, Harvard University, Heidelberg University, The Flatiron Institute, The Johns Hopkins University, L'Ecole polytechnique f\'{e}d\'{e}rale de Lausanne (EPFL), Leibniz-Institut f\"{u}r Astrophysik Potsdam (AIP), Max-Planck-Institut f\"{u}r Astronomie (MPIA Heidelberg), Max-Planck-Institut f\"{u}r Extraterrestrische Physik (MPE), Nanjing University, National Astronomical Observatories of China (NAOC), New Mexico State University, The Ohio State University, Pennsylvania State University, Smithsonian Astrophysical Observatory, Space Telescope Science Institute (STScI), the Stellar Astrophysics Participation Group, Universidad Nacional Aut\'{o}noma de M\'{e}xico, University of Arizona, University of Colorado Boulder, University of Illinois at Urbana-Champaign, University of Toronto, University of Utah, University of Virginia, Yale University, and Yunnan University.  \\

This publication makes use of data products from the Wide-field Infrared Survey Explorer, which is a joint project of the University of California, Los Angeles, and the Jet Propulsion Laboratory/California Institute of Technology, funded by the National Aeronautics and Space Administration. This publication also makes use of data products from NEOWISE, which is a project of the Jet Propulsion Laboratory/California Institute of Technology, funded by the Planetary Science Division of the National Aeronautics and Space Administration.\\

This work has made use of data from the Asteroid Terrestrial-impact Last Alert System (ATLAS) project. 
The Asteroid Terrestrial-impact Last Alert System (ATLAS) project is primarily funded to search for near earth asteroids through NASA grants NN12AR55G, 80NSSC18K0284, and 80NSSC18K1575; byproducts of the NEO search include images and catalogs from the survey area. 
This work was partially funded by Kepler/K2 grant J1944/80NSSC19K0112 and HST GO-15889, and STFC grants ST/T000198/1 and ST/S006109/1. 
The ATLAS science products have been made possible through the contributions of the University of Hawaii Institute for Astronomy, the Queen’s University Belfast, the Space Telescope Science Institute, the South African Astronomical Observatory, and The Millennium Institute of Astrophysics (MAS), Chile.	
This work made use of the data from LAMOST (Large Sky Area Multi-Object Fiber Spectroscopic Telescope, also known as the Guoshoujing Telescope) (https://cstr.cn/31118.02.LAMOST). LAMOST is a Chinese national mega-science facility, operated by National Astronomical Observatories, Chinese Academy of Sciences.\\

This work made use of data from the Catalina Real-time Transient Survey, which is based on data obtained by the Catalina Sky Survey. The Catalina Sky Survey is funded by NASA under the Near Earth Object Observations Program.\\

This research is supported by the National Natural Science Foundation of
China (12273033) and the Xingdian Talent Support Project of
the Yunnan Province (XDYC-YLXZ-2023-0016)
	
\end{acknowledgements}

\bibliography{cl3}
\bibliographystyle{aa}

\begin{appendix}
\onecolumn

	\section{Information for the DOT and HCT spectroscopic observations of the 12 candidates}
	\begin{table*}[htbp]

\centering

\caption{Information for the DOT and HCT Observations}

\label{tab:obs}

\begin{tabular}{lcccc}

\hline

Name & R.A. & Dec. & Standard & Observation \\

\hline

\multicolumn{5}{l}{\textbf{Confirmed CL AGN}} \\

\hline

J014638+131109 & 26.6620   & 13.1859   & G19B2B  & 2024-12-23/HCT \\

J115953+051330 & 179.9725  & 5.2250    & Feige34 & 2025-01-25/DOT \\

J121228+501412 & 183.1187  & 50.2367   & Feige66 & 2025-03-24/HCT \\

J131101+000310 & 197.7551  & 0.0530    & Feige34 & 2025-01-24/DOT \\

J134330+510204 & 205.8751  & 51.0345   & Feige34 & 2025-01-23/DOT \\

J152502+110744 & 231.2622  & 11.1289   & HZ44    & 2025-05-28/HCT  \\

J160626+044802 & 241.6099  & 4.8006    & Feige66 & 2025-03-24/HCT \\

\hline

\multicolumn{5}{l}{\textbf{Other candidate}} \\

\hline

J075846+270515 & 119.6958 & 27.0877 & G19B2B  & 2024-12-23/HCT \\

J111614+061736 & 169.0611 & 6.29335   & Feige66 & 2025-03-24/HCT \\

J105344+492955 & 163.4339 & 49.4989 & Feige34 & 2025-01-21/DOT \\

J111137+140118 & 167.9066 & 14.0217   & Feige34 & 2025-01-23/DOT \\

J132558+411500 & 201.4956 & 41.2501 & HZ44    & 2025-05-28/HCT \\

\hline

\end{tabular}

\end{table*}

\clearpage

	\section{Light curves and spectra of seven CL AGNs}
\begin{figure*}[htbp]
	\centering
	\includegraphics[width=0.49\textwidth]{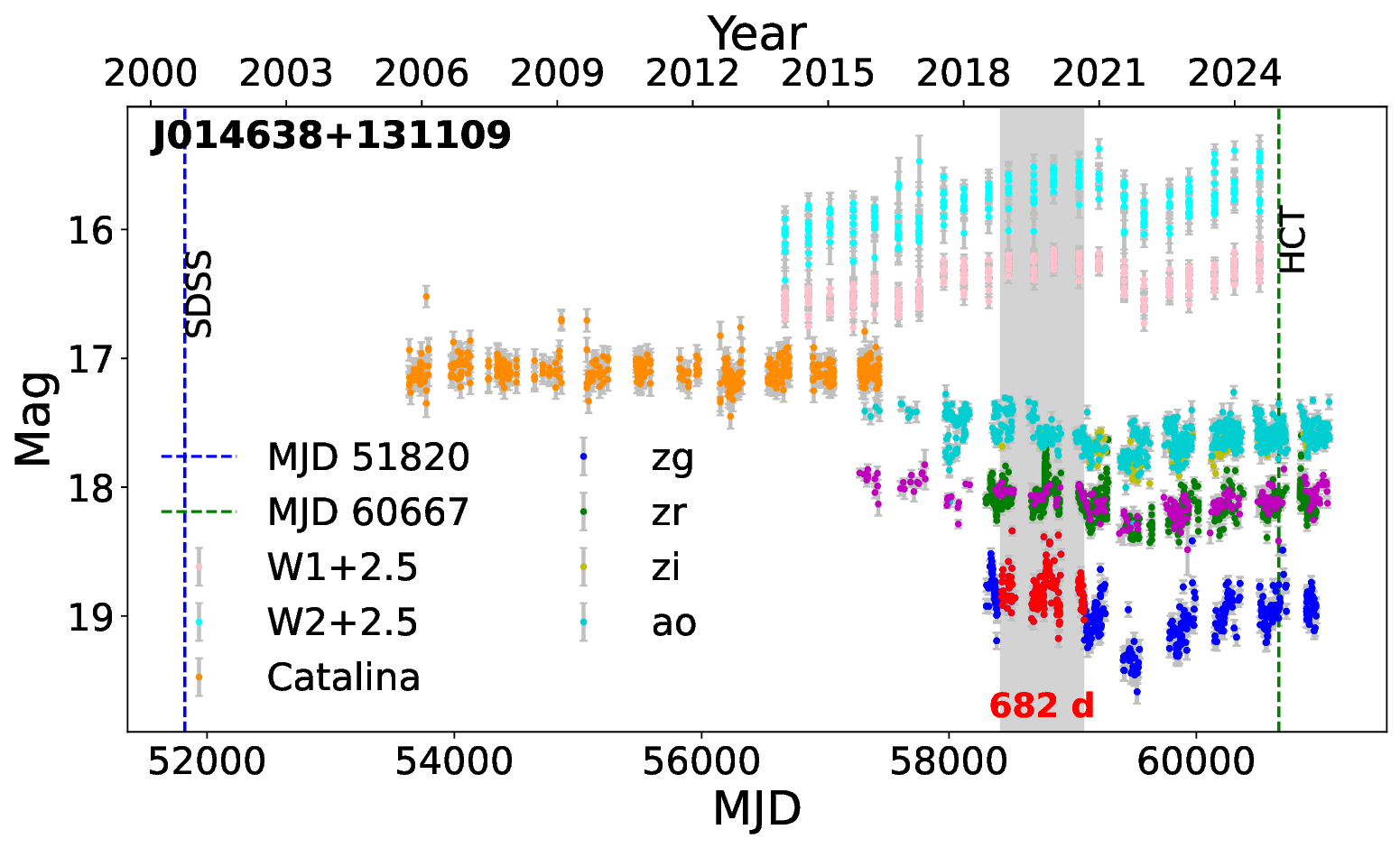}
	\hfill
	\includegraphics[width=0.49\textwidth]{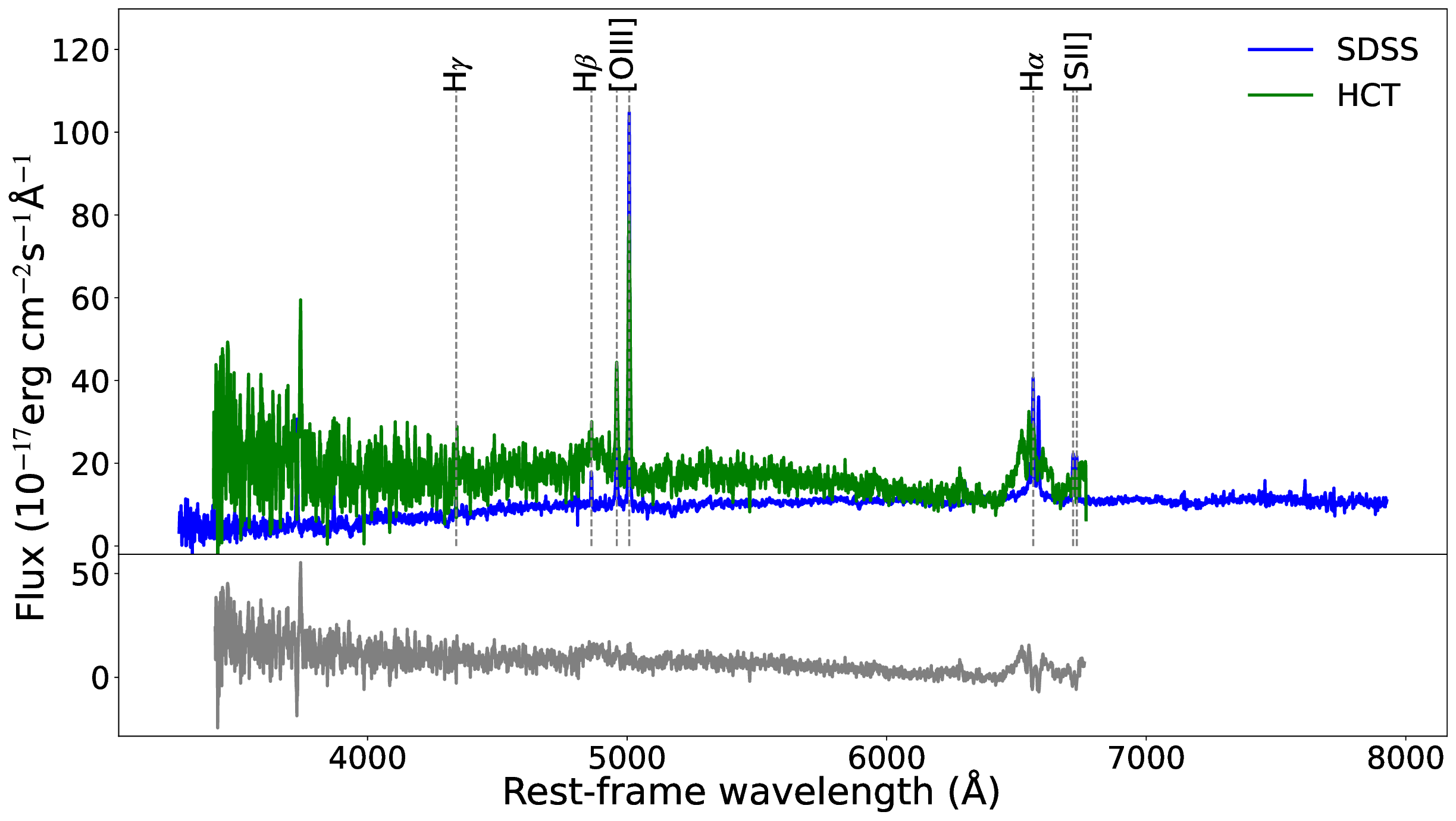}
	\caption{Same as Fig.~\ref{fig:J1606} for J014638+131109.}
	\label{fig:J0146}
\end{figure*}


\begin{figure*}[htbp]
	\centering
	\includegraphics[width=0.49\textwidth]{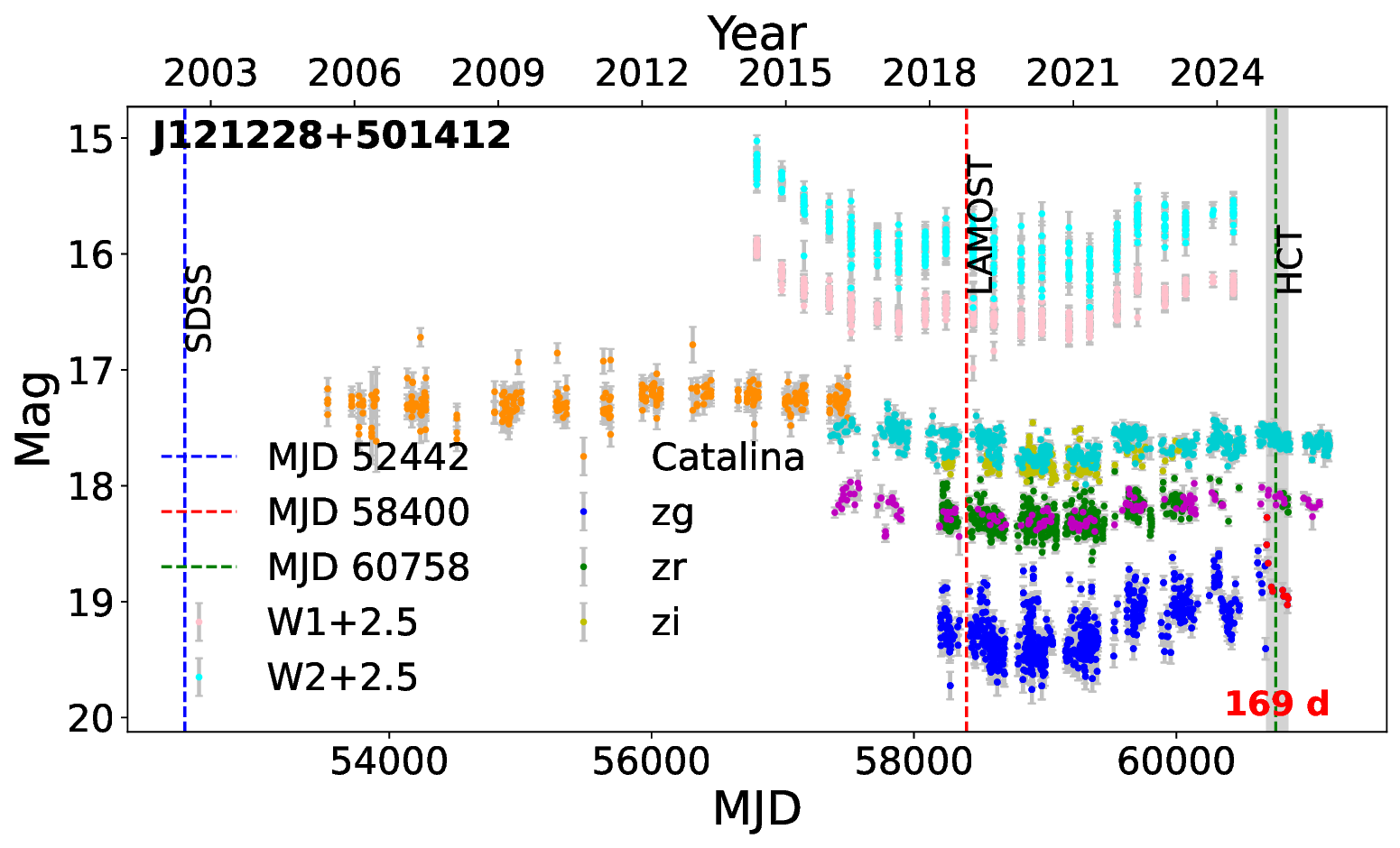}
	\hfill
	\includegraphics[width=0.49\textwidth]{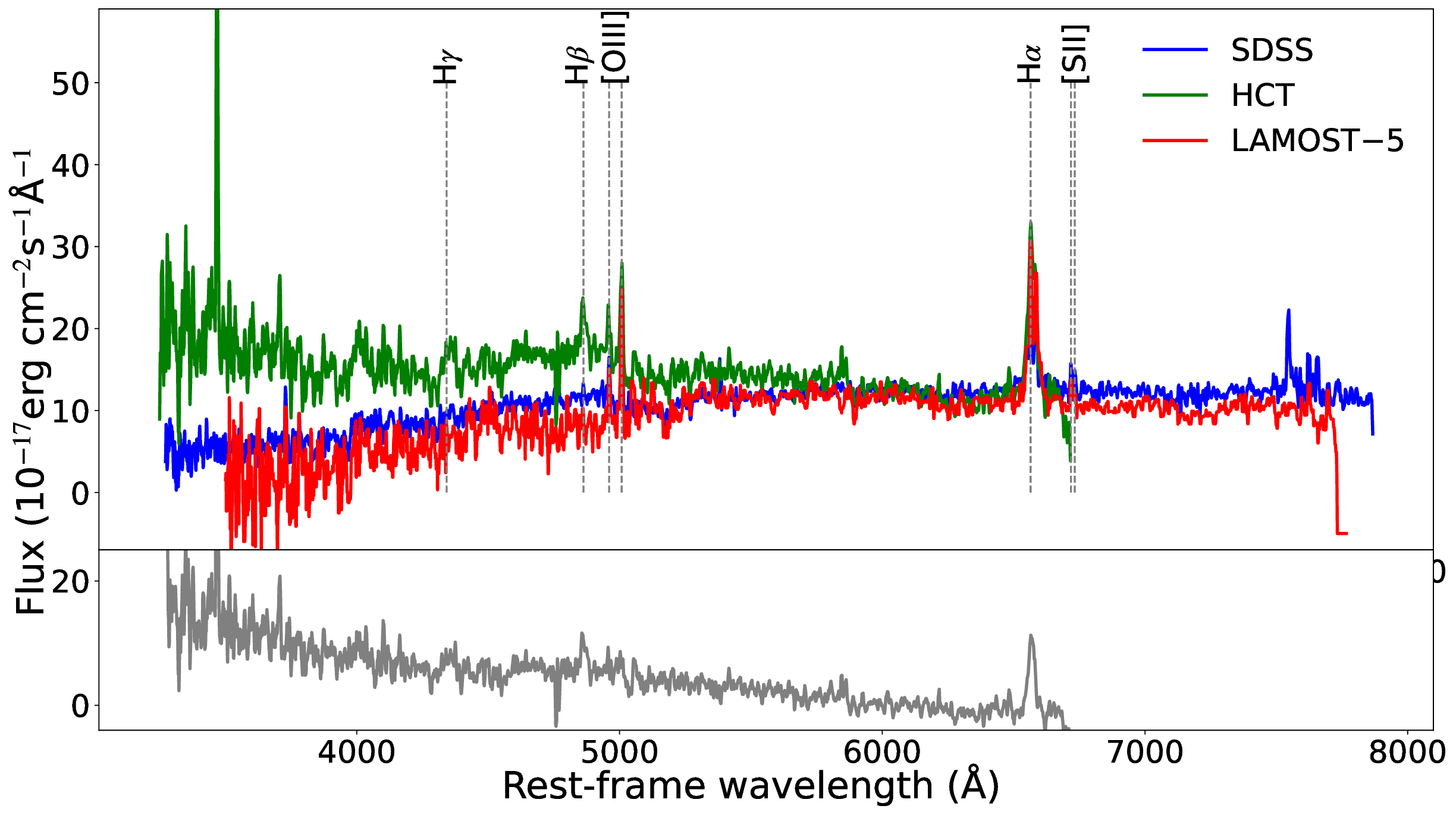}
	\caption{Same as Fig.~\ref{fig:J1606} for J121228+501412.}\label{fig:J1212}
\end{figure*}

\begin{figure*}[htbp]
	\centering
	\includegraphics[width=0.49\textwidth]{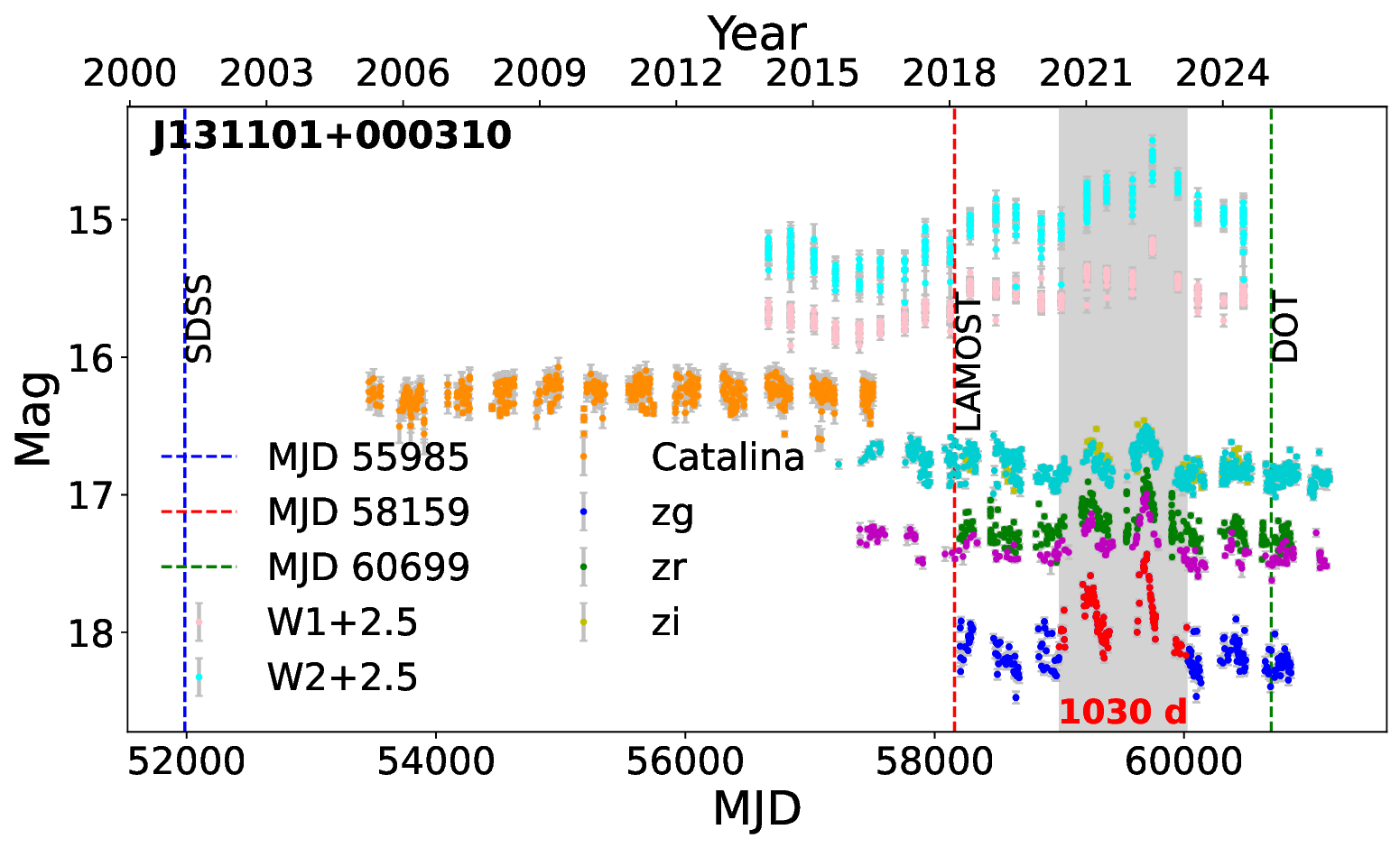}
	\hfill
	\includegraphics[width=0.49\textwidth]{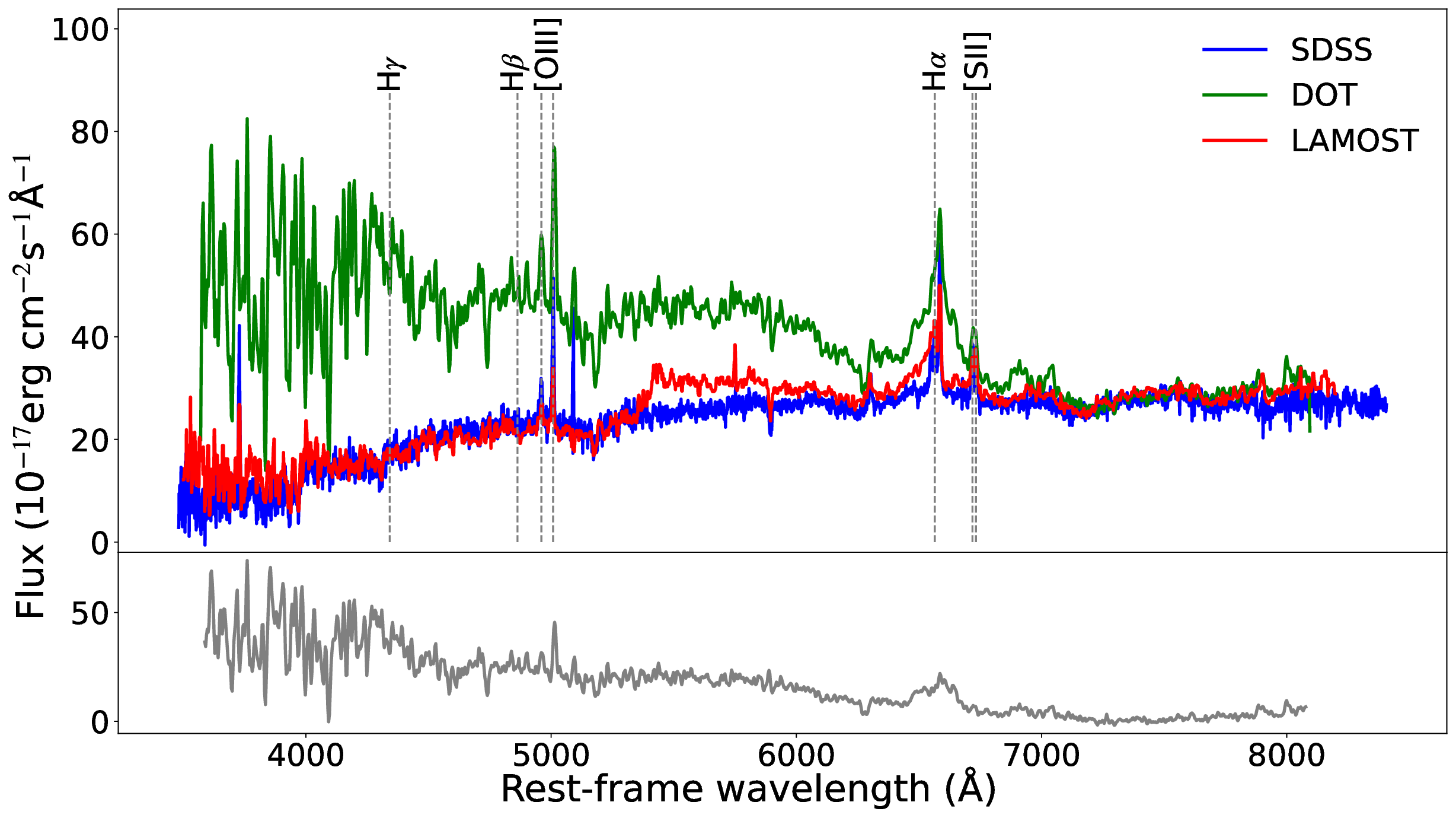}
	\caption{Same as Fig.~\ref{fig:J1159} for J131101+000310.}\label{fig:J1311}
\end{figure*}

\begin{figure*}[htbp]
	\centering
	\includegraphics[width=0.49\textwidth]{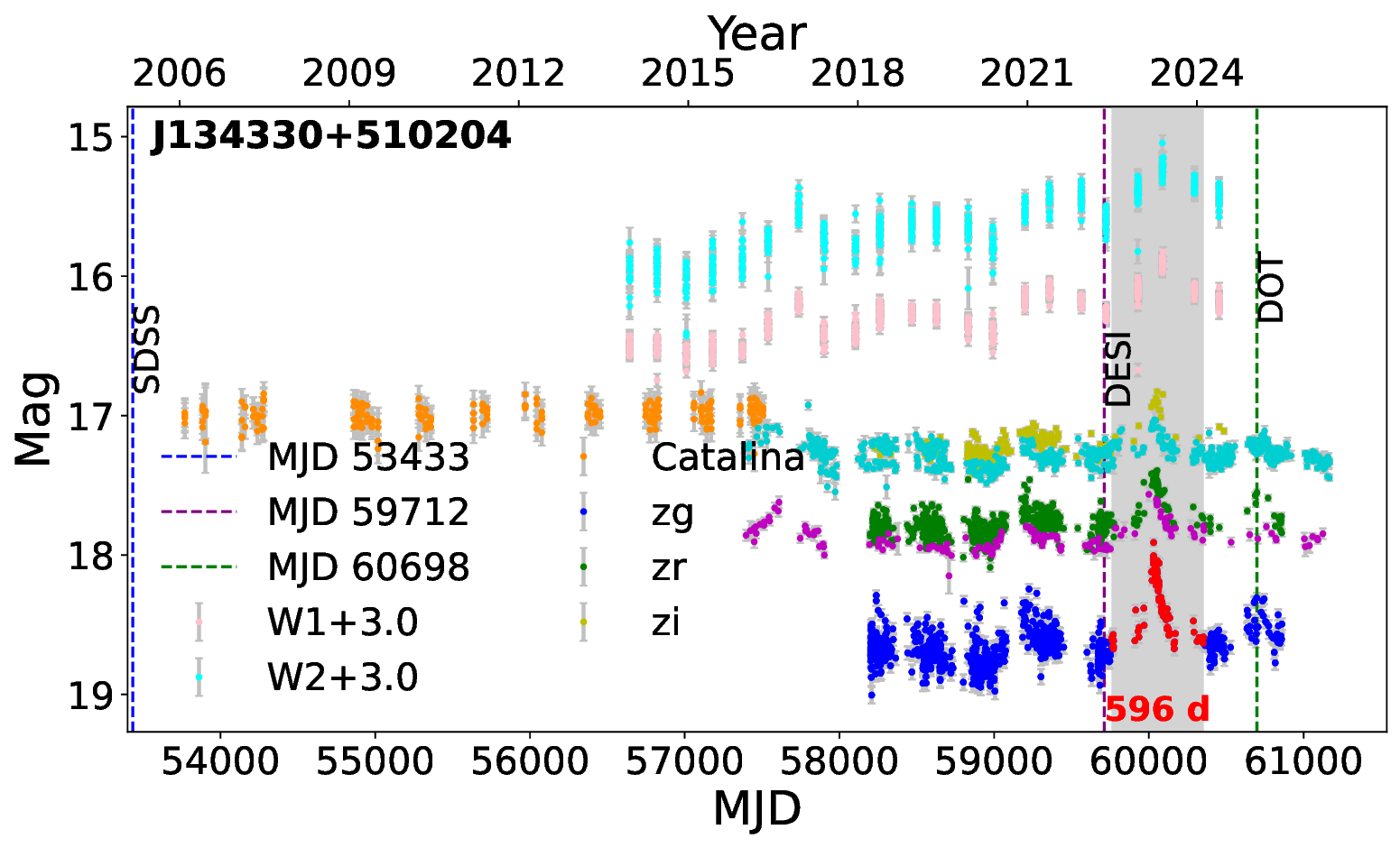}
	\hfill
	\includegraphics[width=0.49\textwidth]{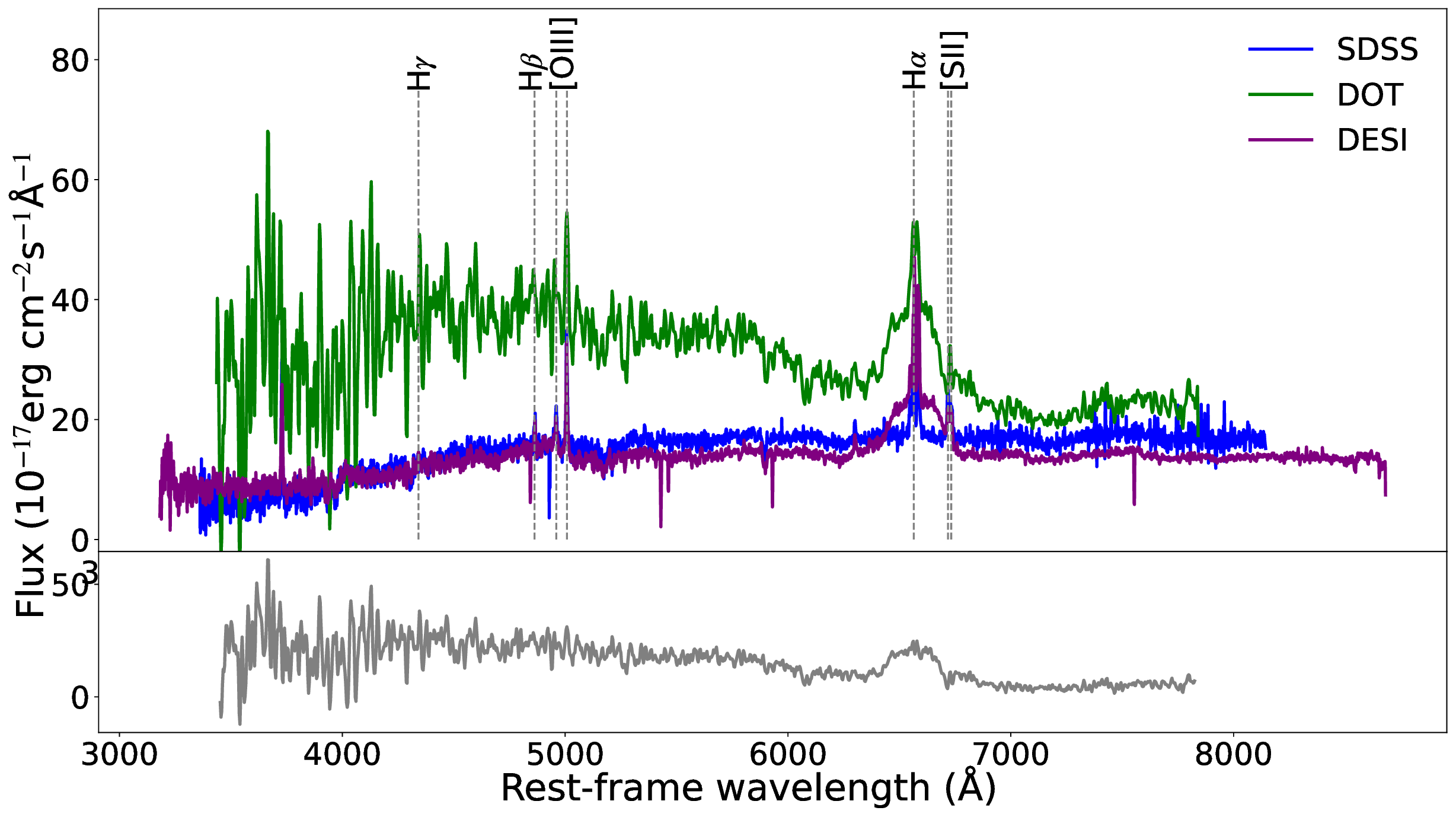}
	\caption{Same as Fig.~\ref{fig:J1159} for J134330+510204.}\label{fig:J1343}
\end{figure*}

\begin{figure*}[htbp]
	\centering
	\includegraphics[width=0.49\textwidth]{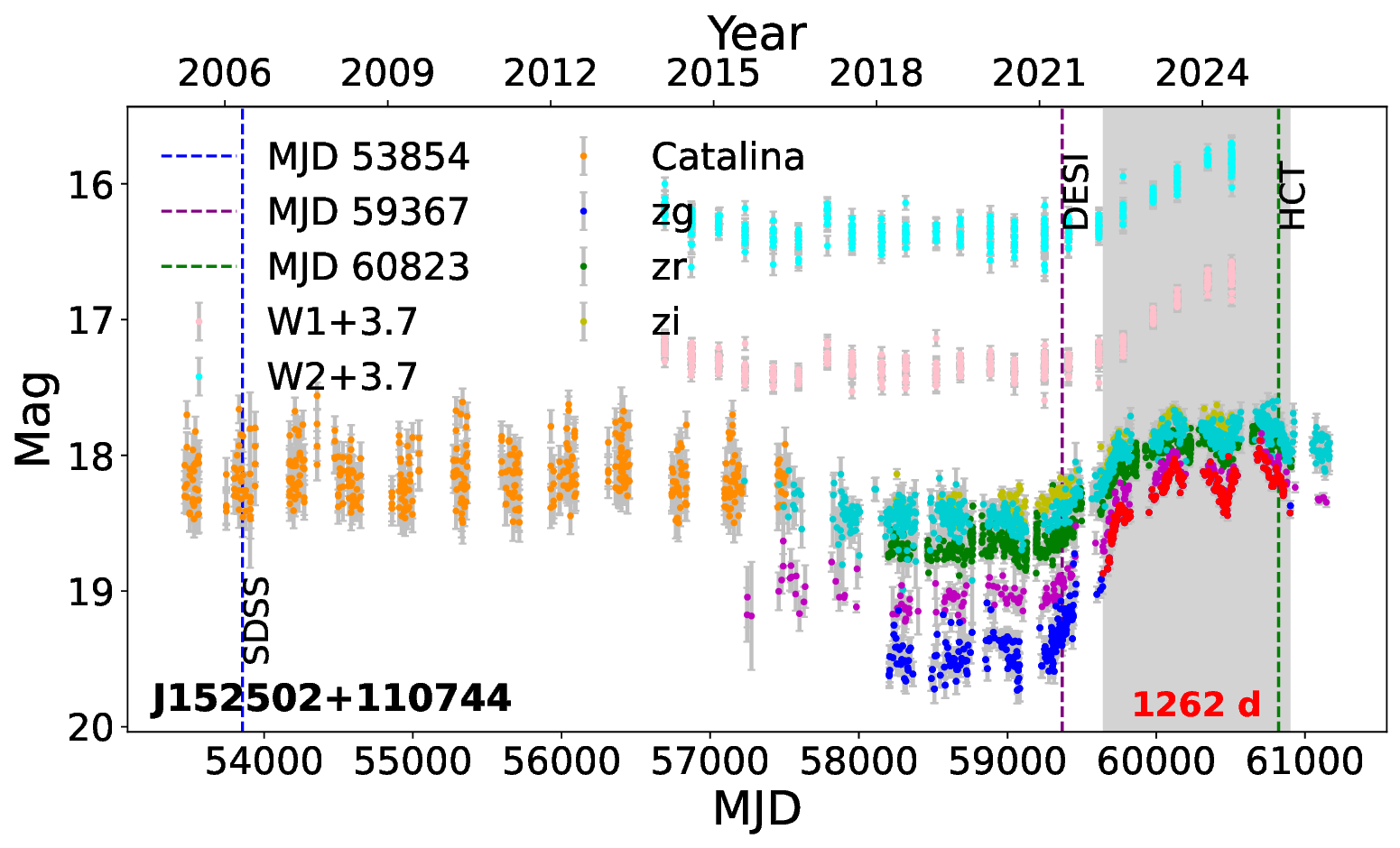}
	\hfill
	\includegraphics[width=0.49\textwidth]{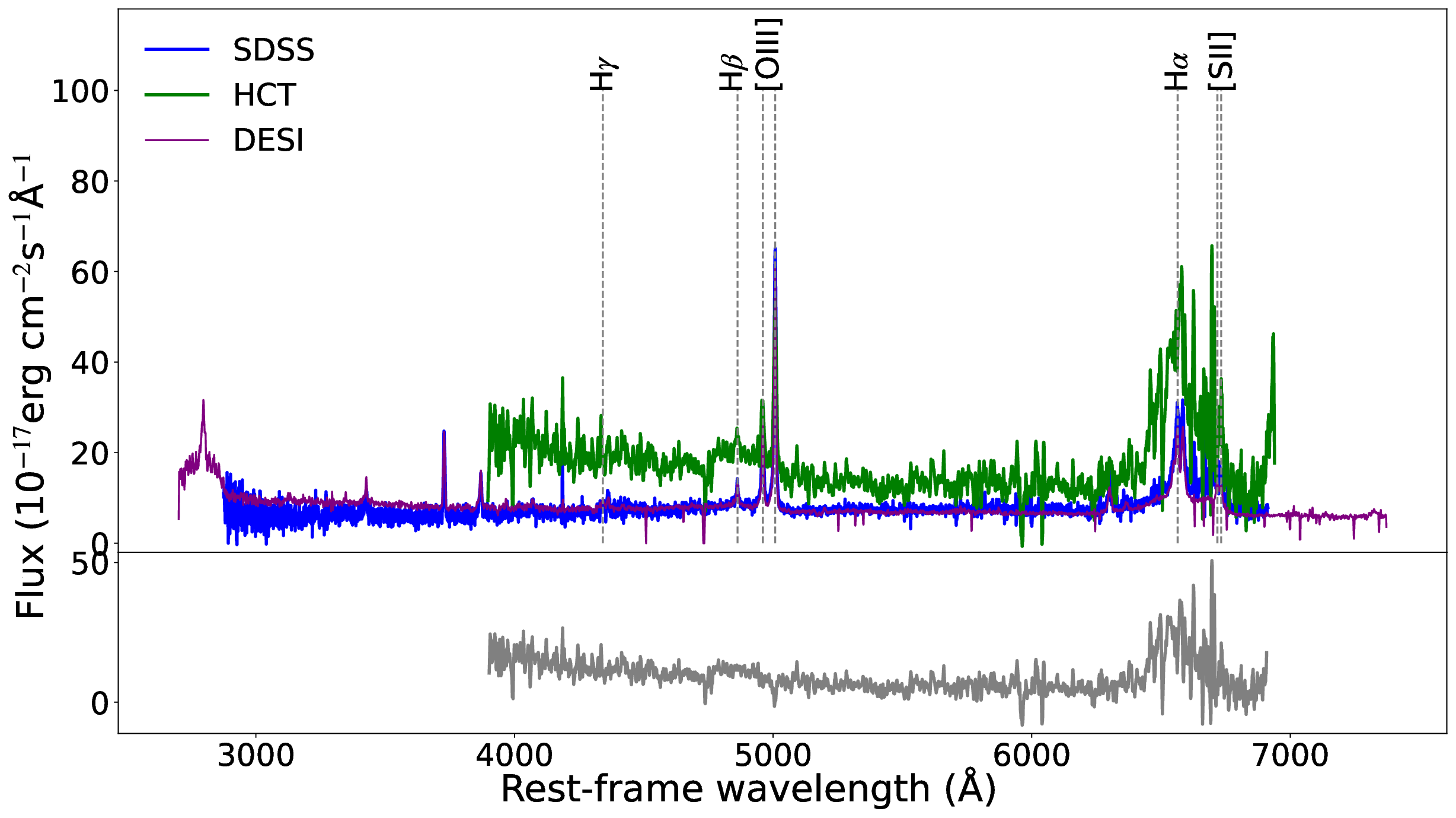}
	\caption{Same as Fig.~\ref{fig:J1606} for J152502+110744.}\label{fig:J1525}
\end{figure*}

\clearpage

\section{Light curves and spectra of five other candidates}

\begin{figure*}[htbp]
	\centering
	\includegraphics[width=0.49\textwidth]{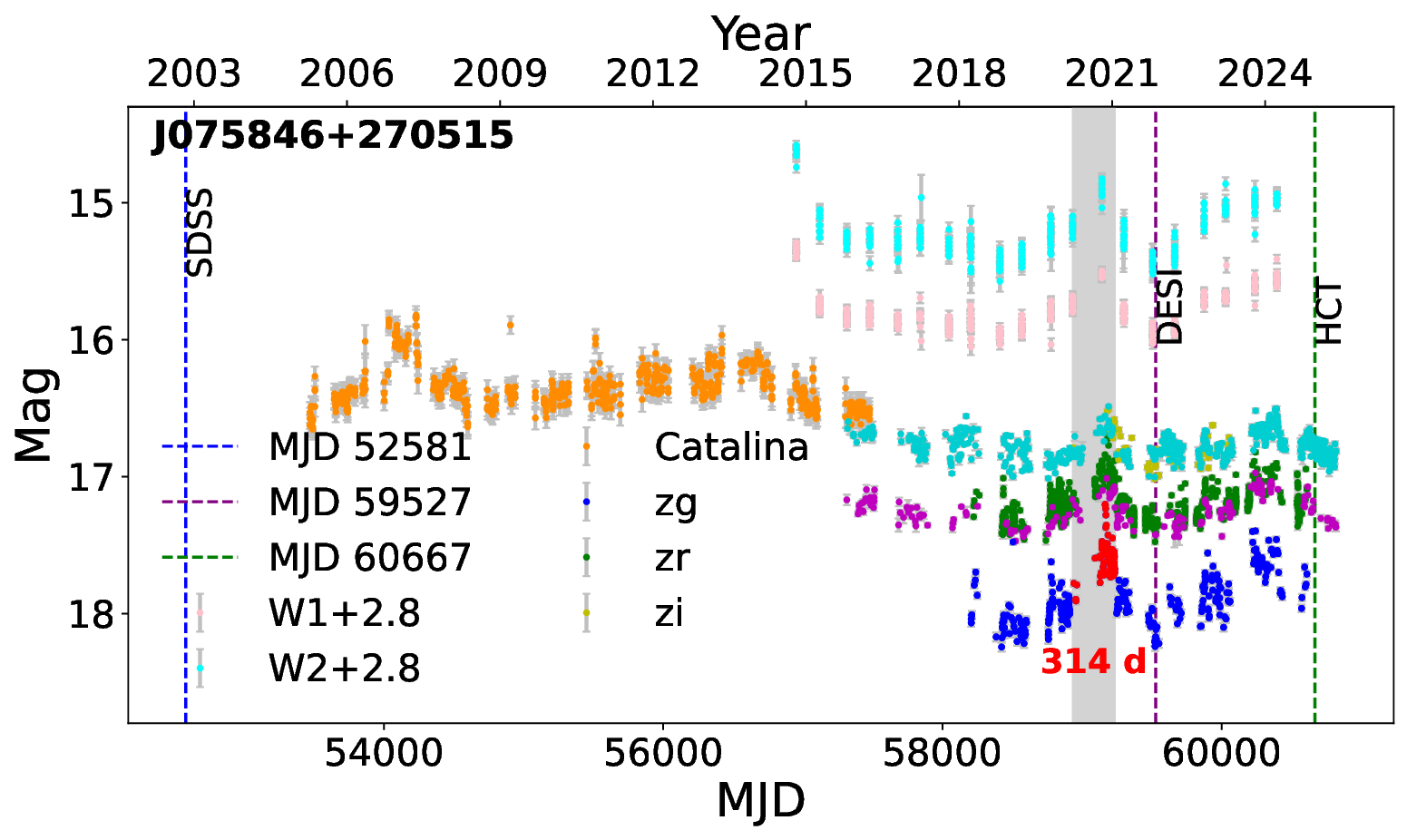}
	\hfill
	\includegraphics[width=0.49\textwidth]{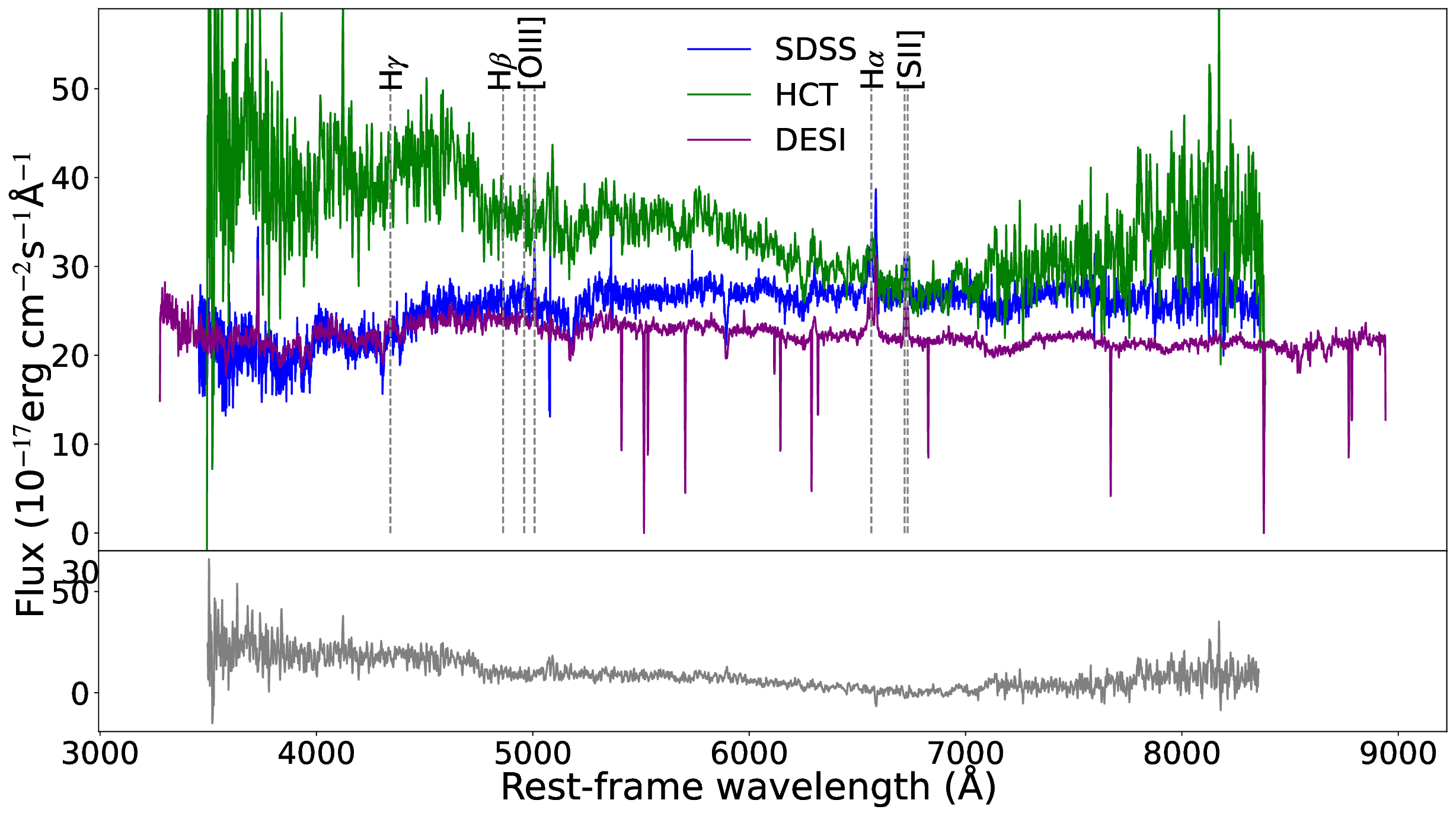}
	\caption{Same as Fig.~\ref{fig:J1606} for not confirmed CL candidate J075846+270515.}\label{fig:J0758}
\end{figure*}

\begin{figure*}[htbp]
	\centering
	\includegraphics[width=0.49\textwidth]{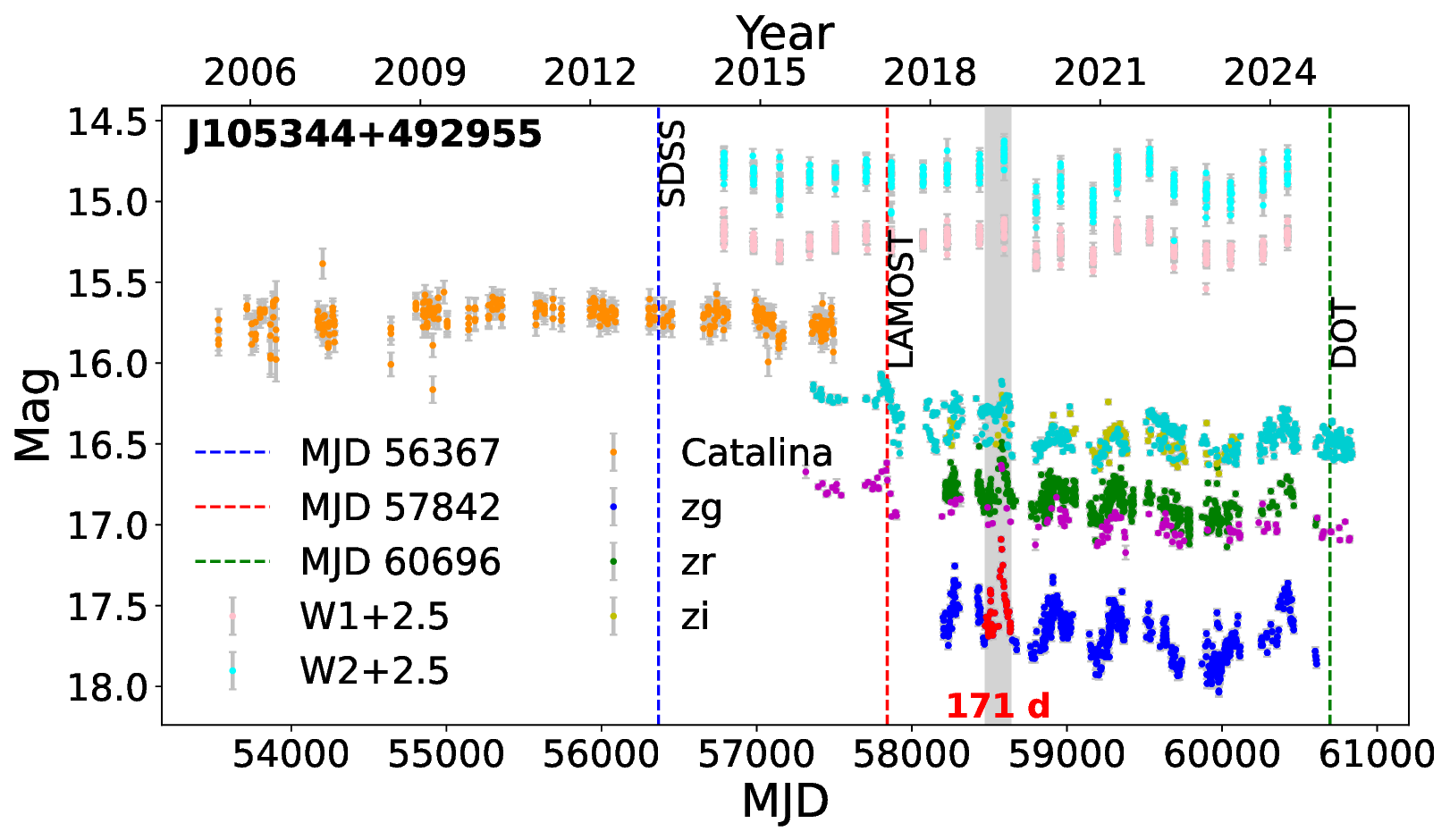}
	\hfill
	\includegraphics[width=0.49\textwidth]{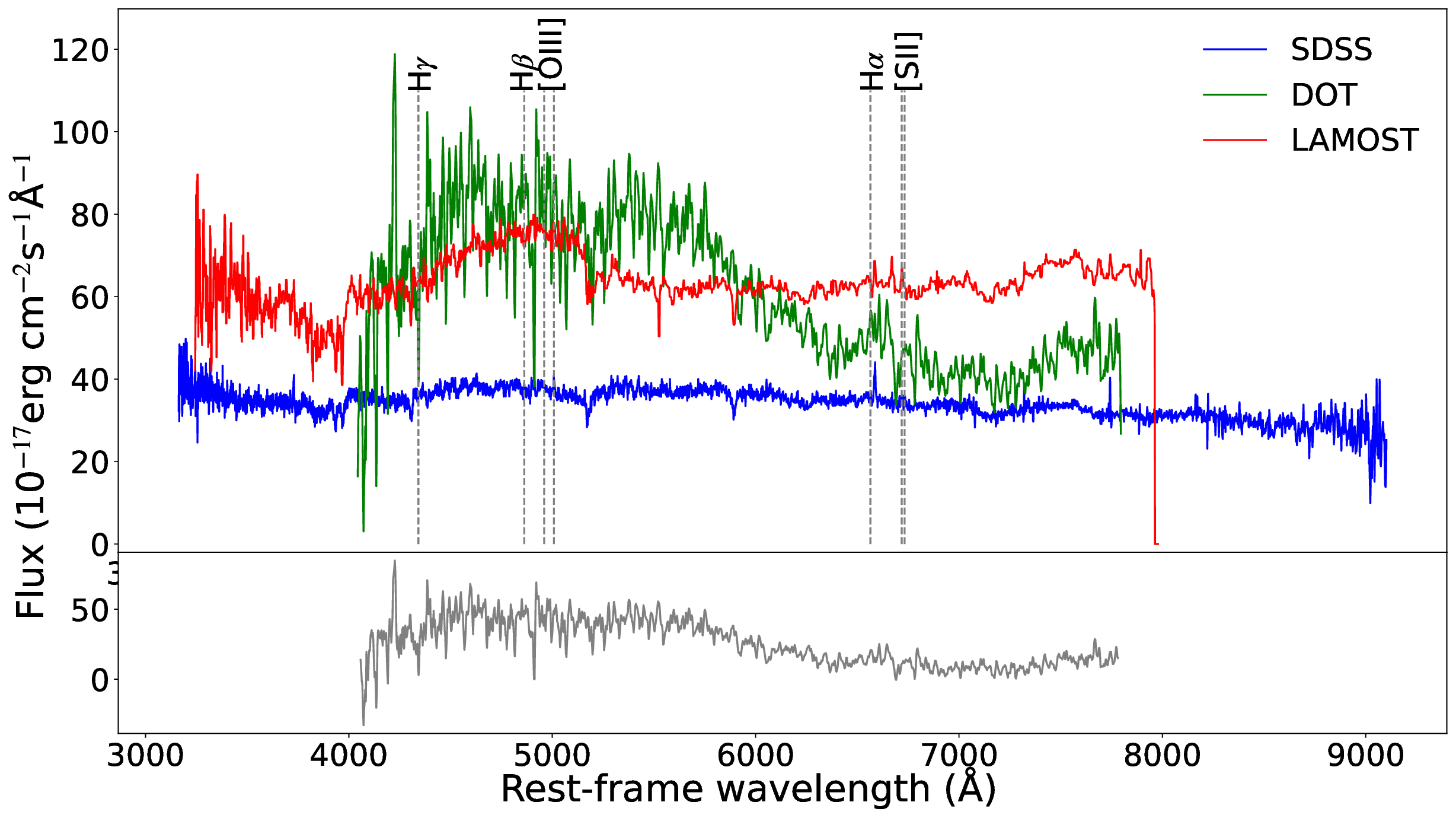}
	\caption{Same as Fig.~\ref{fig:J1159} for not confirmed CL candidate J105344+492955.}\label{fig:J1053}
\end{figure*}

\begin{figure*}[htbp]
	\centering
	\includegraphics[width=0.49\textwidth]{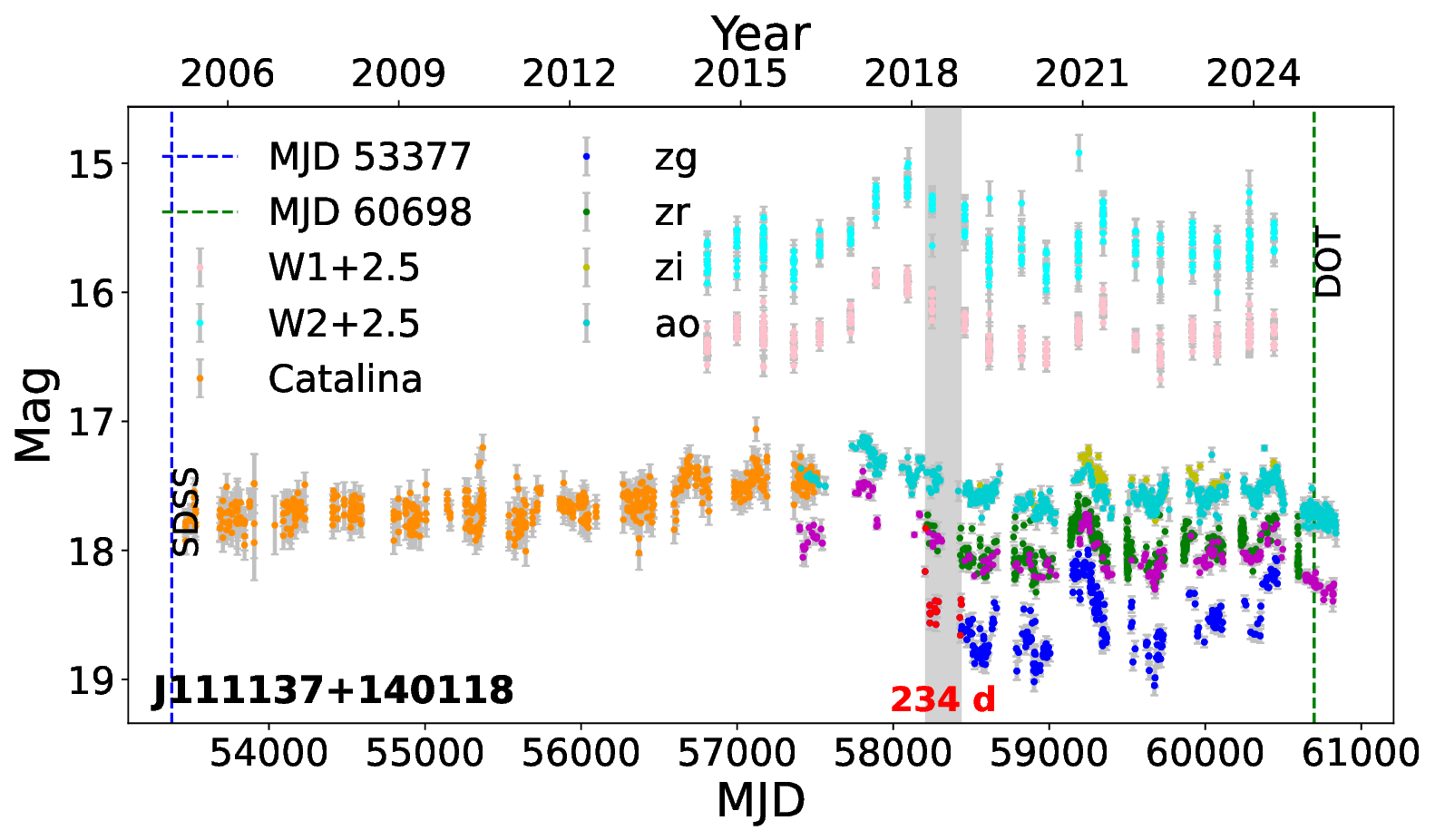}
	\hfill
	\includegraphics[width=0.49\textwidth]{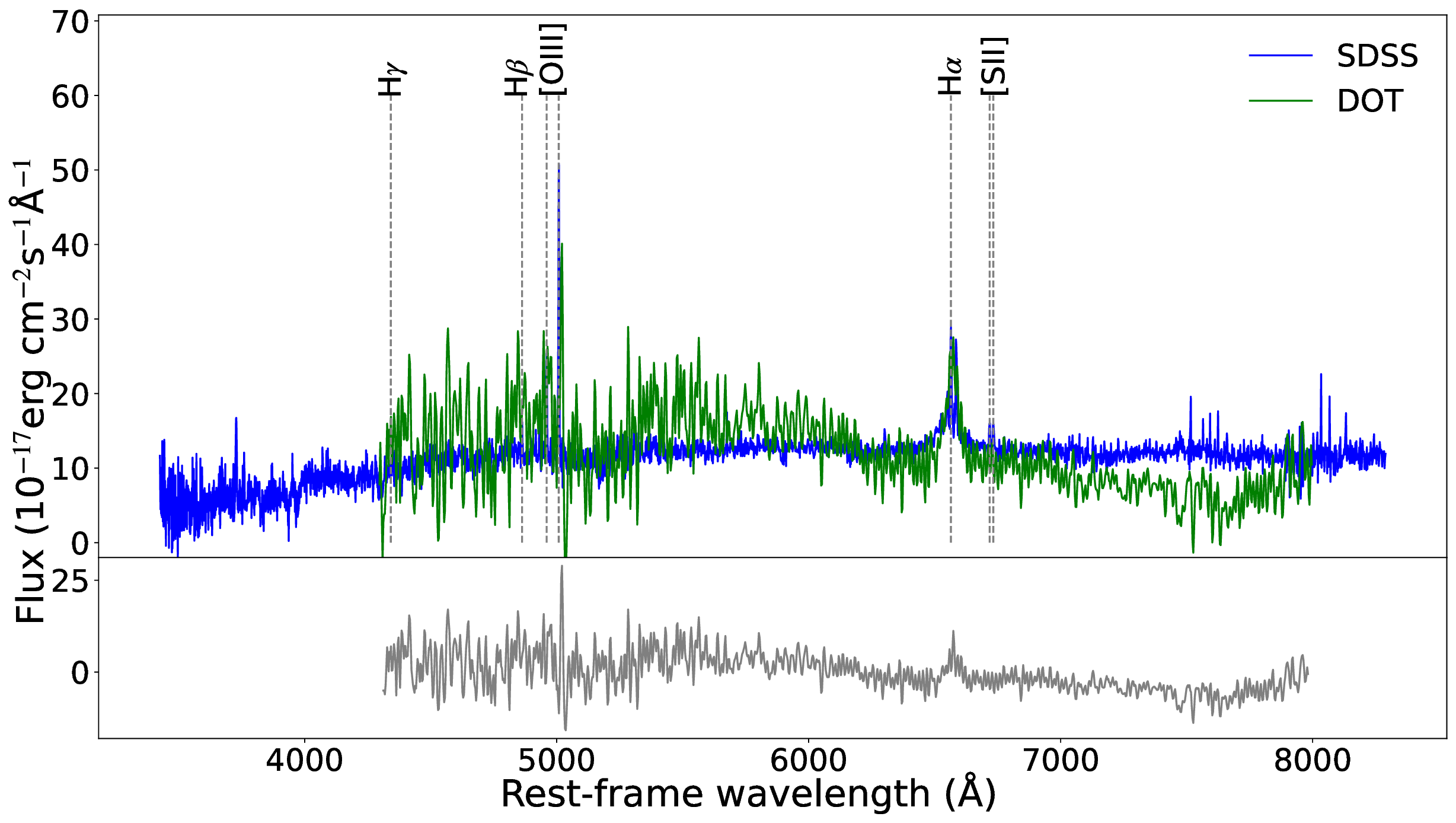}
	\caption{Same as Fig.~\ref{fig:J1159} for not confirmed CL candidate J111137+140118.}\label{fig:J1111}
\end{figure*}

\begin{figure*}[htbp]
	\centering
	\includegraphics[width=0.49\textwidth]{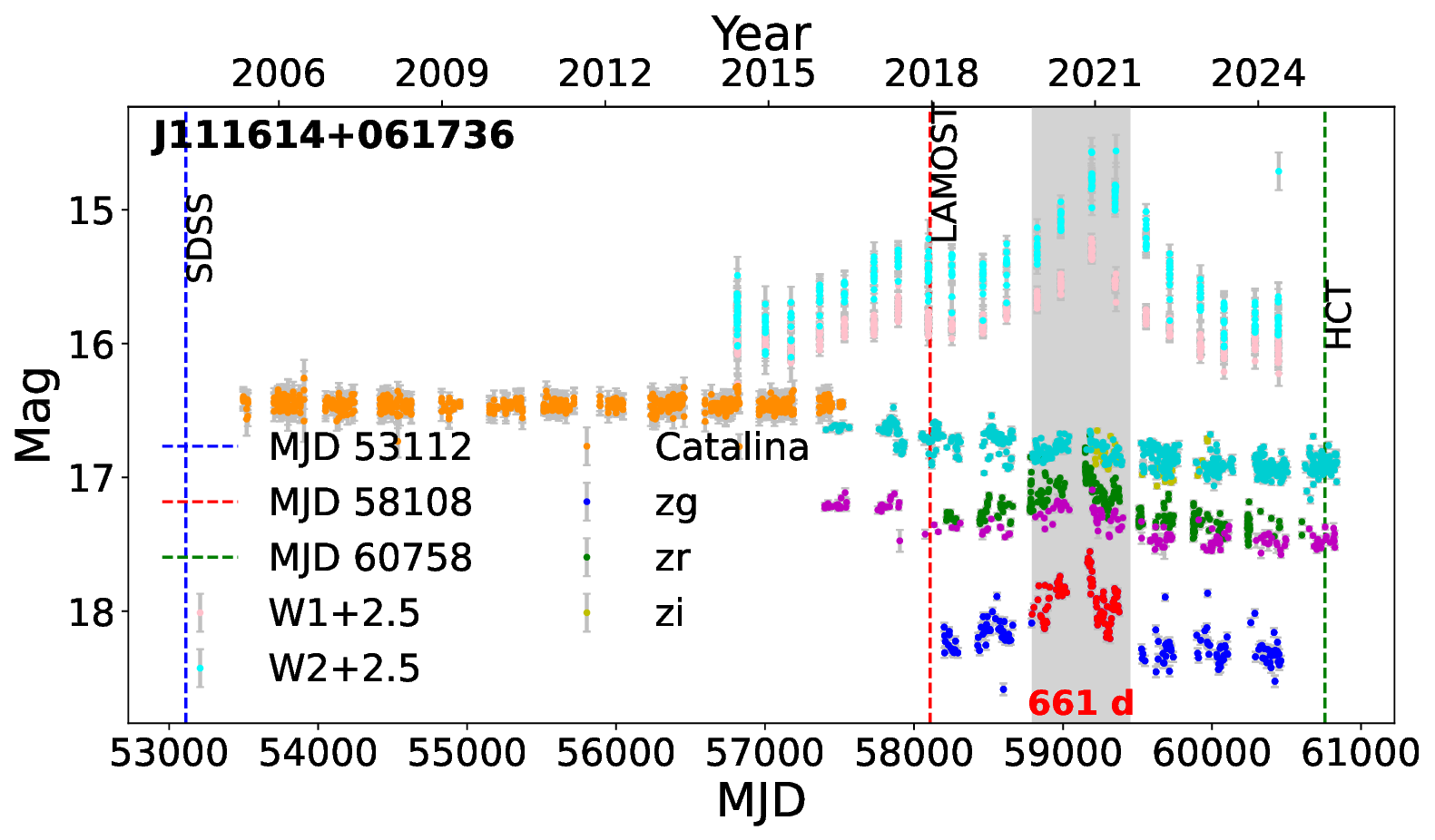}
	\hfill
	\includegraphics[width=0.49\textwidth]{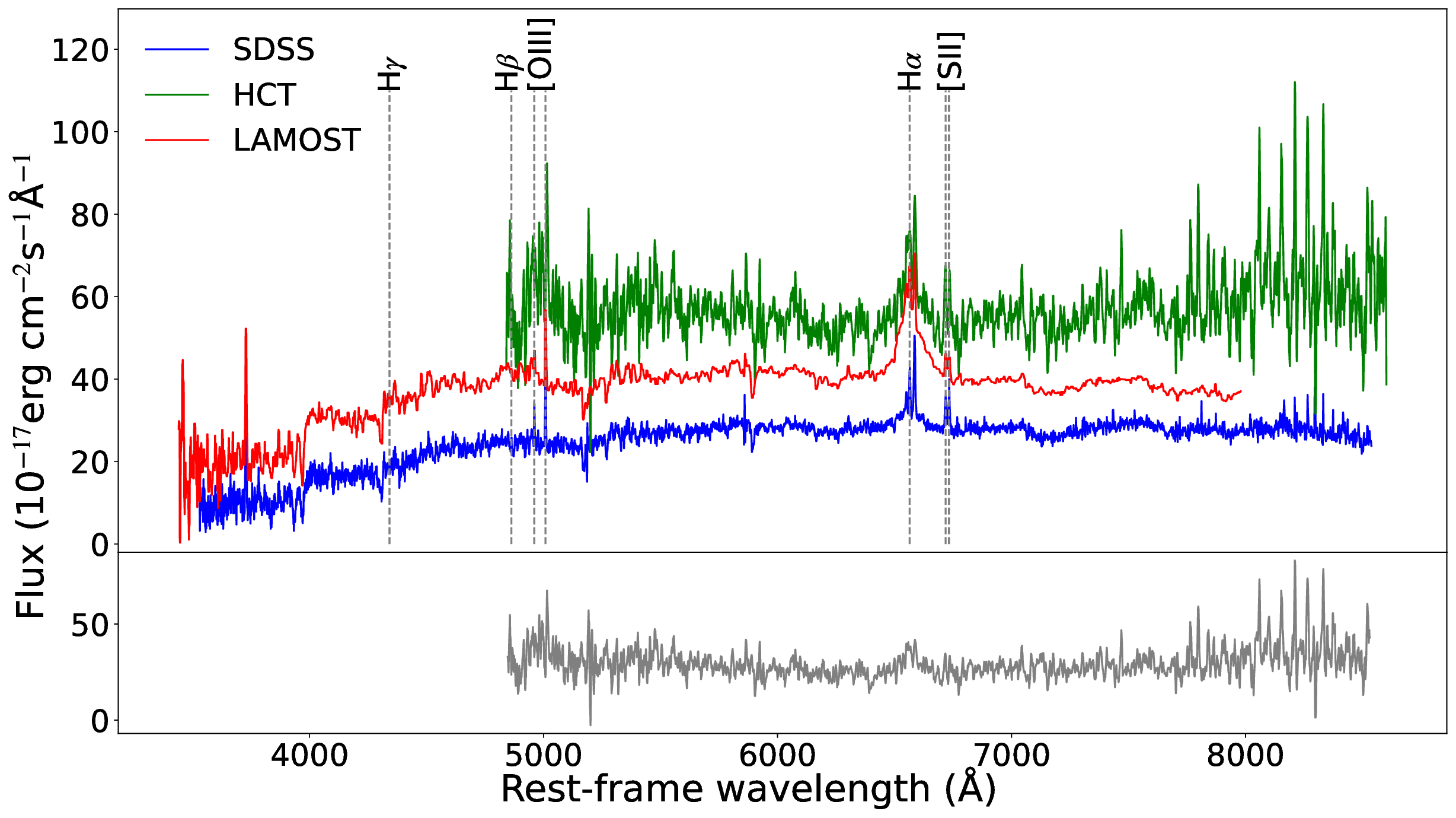}
	\caption{Same as Fig.~\ref{fig:J1606} for not confirmed CL candidate J111614+061736.}\label{fig:J1116}
\end{figure*}

\begin{figure*}[htbp]
	\centering
	\includegraphics[width=0.49\textwidth]{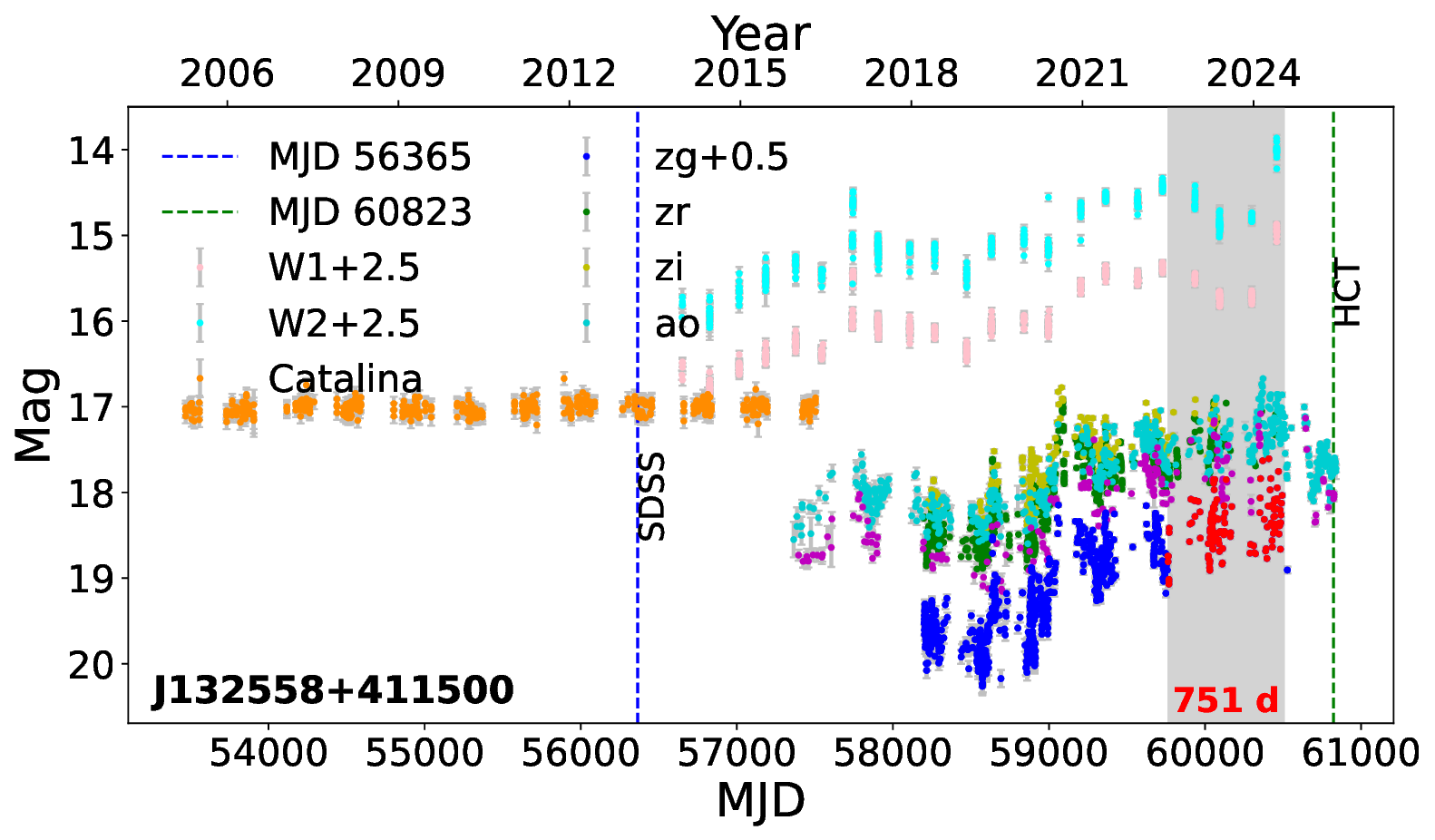}
	\hfill
	\includegraphics[width=0.49\textwidth]{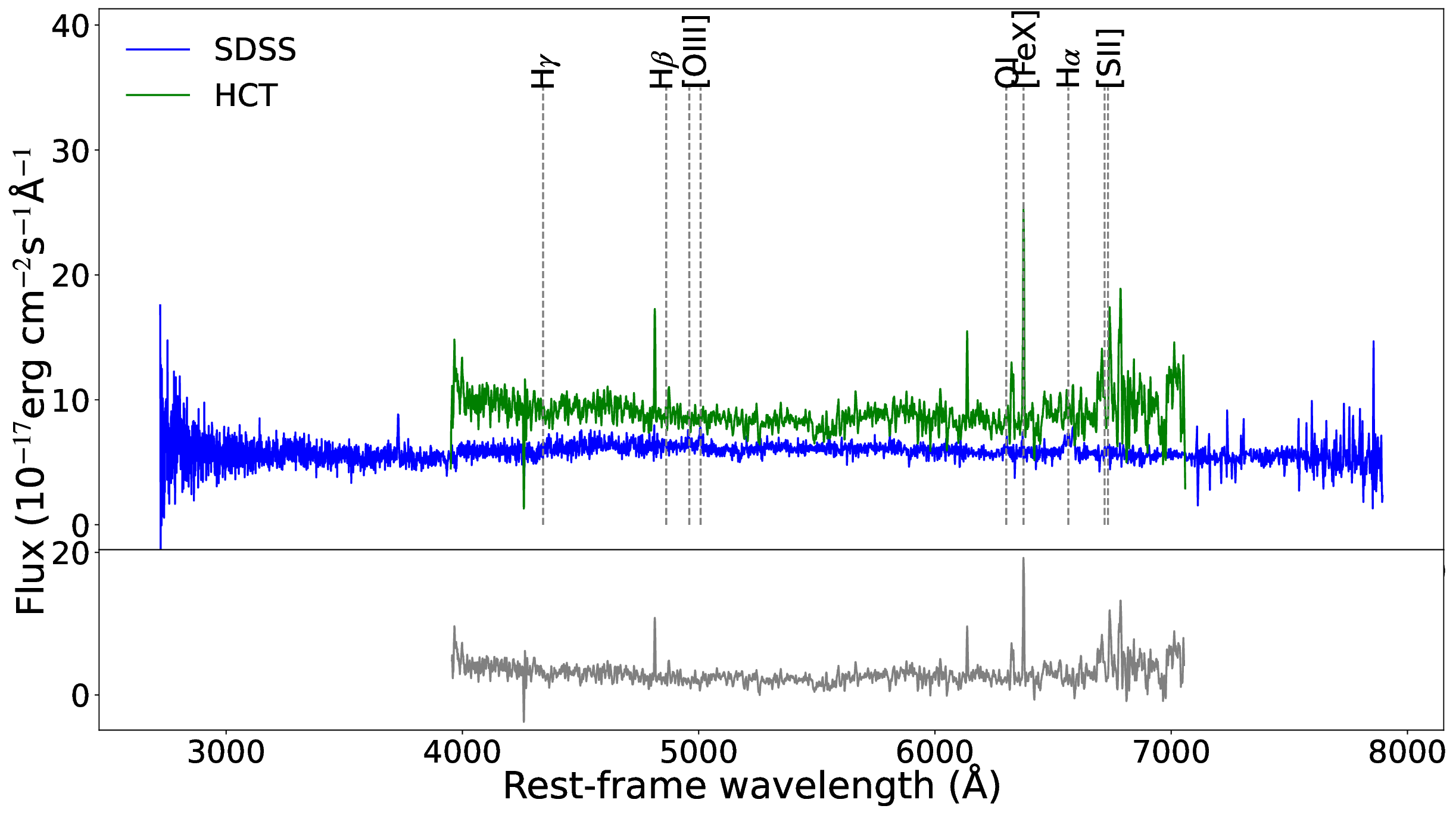}
	\caption{Same as Fig.~\ref{fig:J1606} for not confirmed CL candidate J132558+411500.}\label{fig:J1325}
\end{figure*}

\clearpage

\section{CM diagrams of 12 AGN targets}

\begin{figure*}
	\centering
	\includegraphics[width=0.49\textwidth]{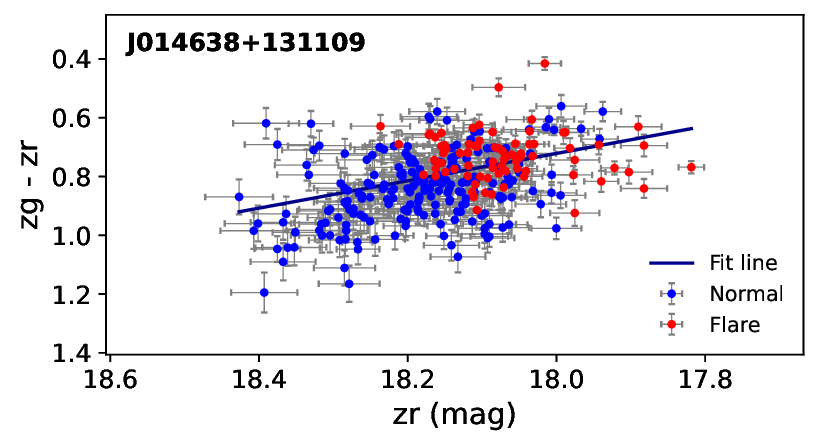}
        \hfill
        \includegraphics[width=0.49\textwidth]{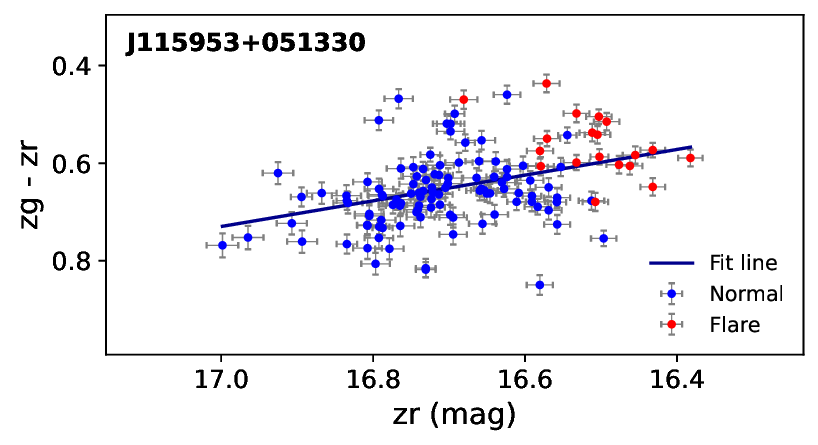}
	\includegraphics[width=0.49\textwidth]{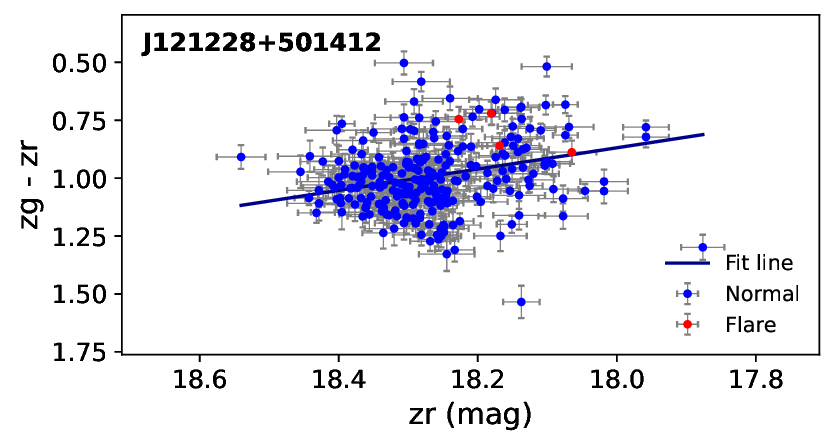}
        \hfill
        \includegraphics[width=0.49\textwidth]{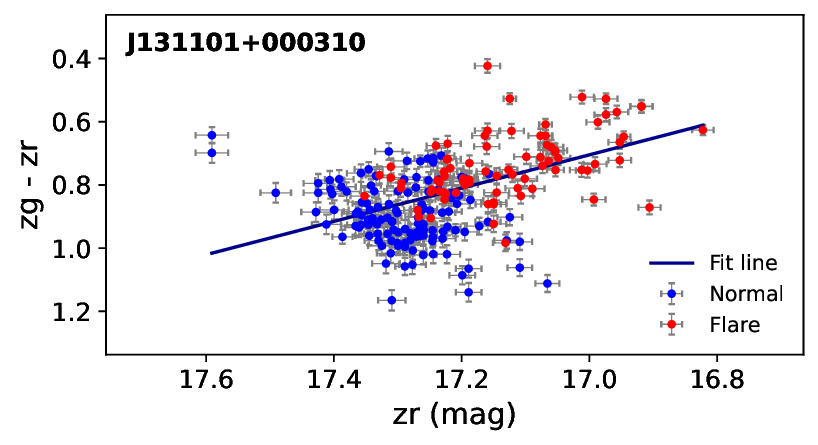}
	\includegraphics[width=0.49\textwidth]{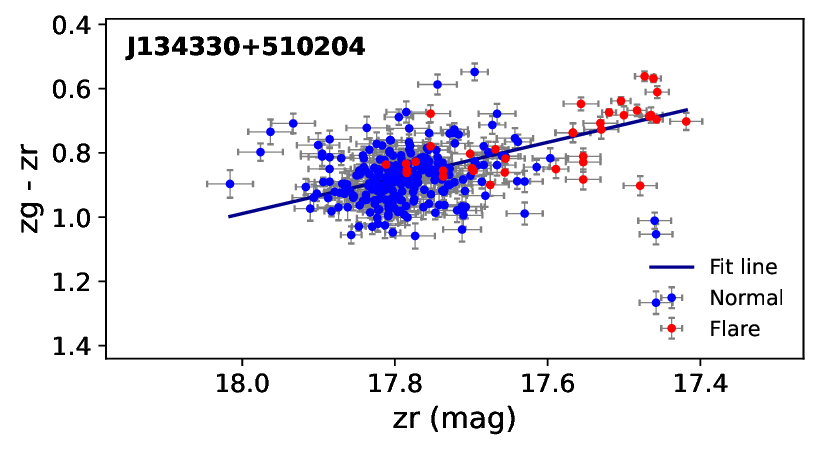}
        \hfill
        \includegraphics[width=0.49\textwidth]{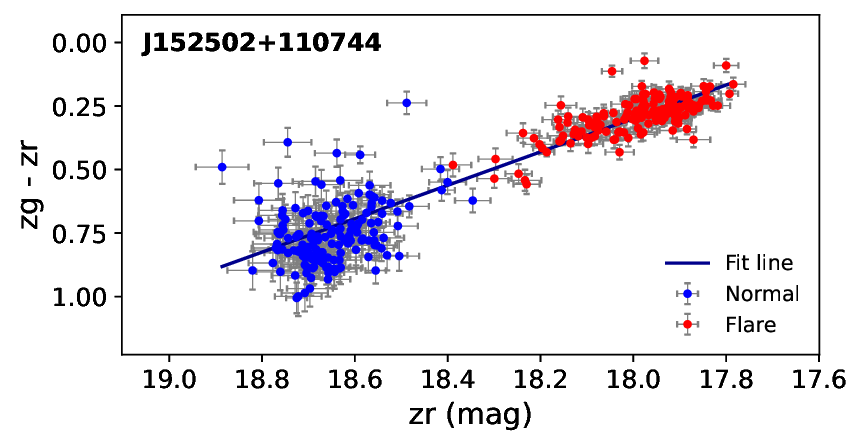}
        \hfill
        \includegraphics[width=0.49\textwidth]{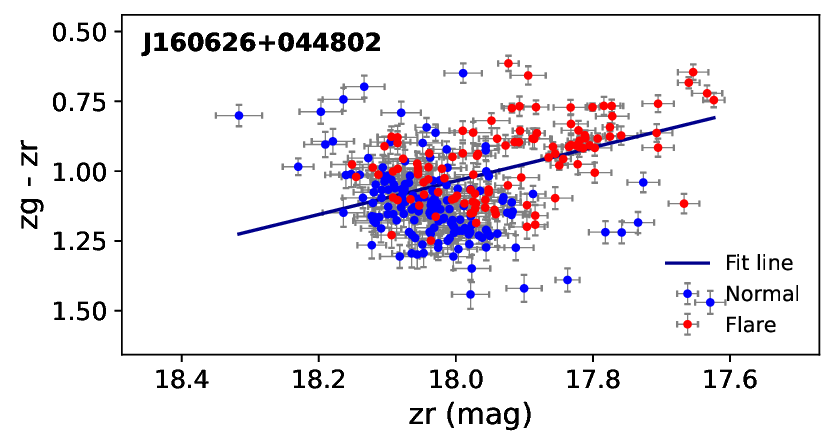}
\caption{CM diagrams of seven CL-AGNs: J014638+131109, J115953+051330, J121228+501412, J131101+000310, J134330+510204, J152502+110744, and J160626+044802.}
	\label{fig:clcm}
\end{figure*}

\begin{figure*}
        \centering
        \includegraphics[width=0.49\textwidth]{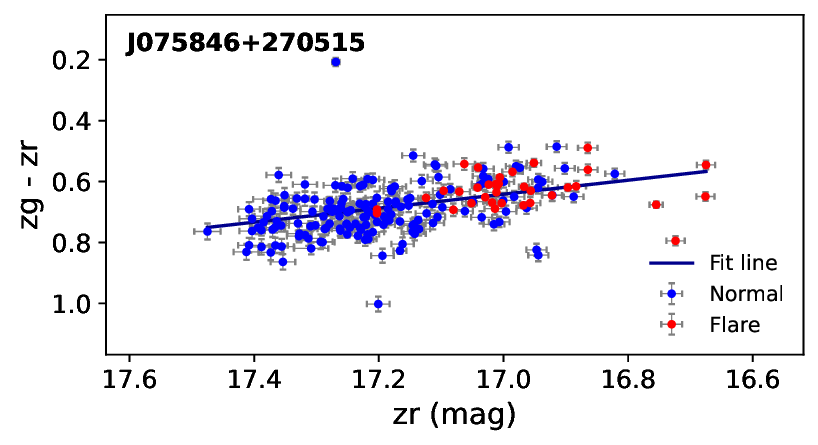}
        \hfill
        \includegraphics[width=0.49\textwidth]{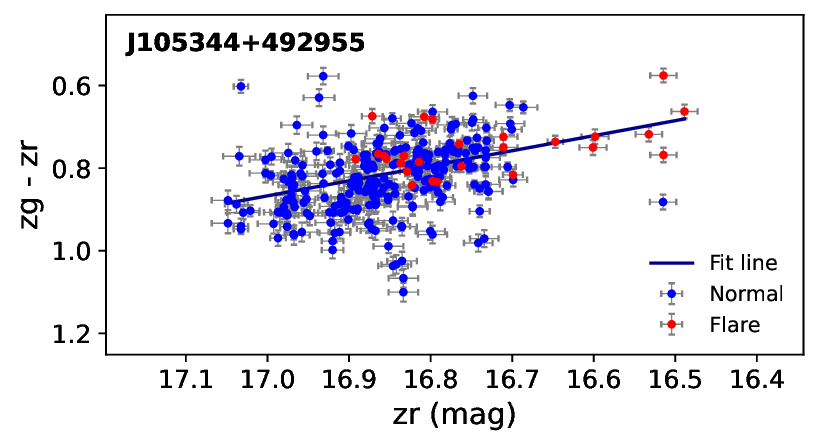}
        \includegraphics[width=0.49\textwidth]{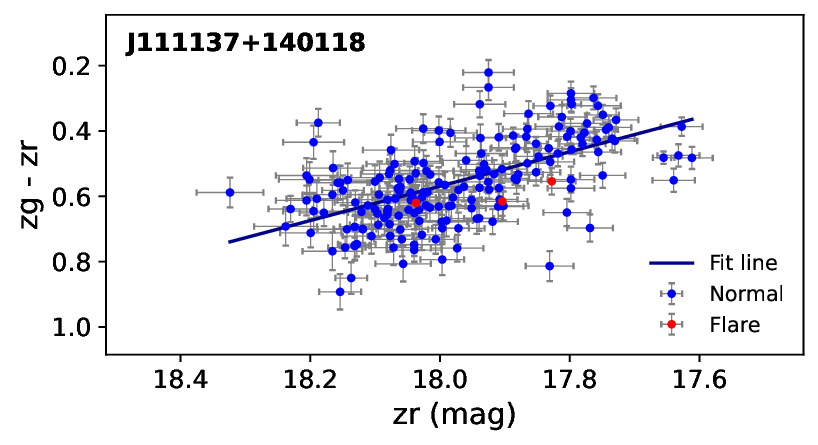}
        \hfill
        \includegraphics[width=0.49\textwidth]{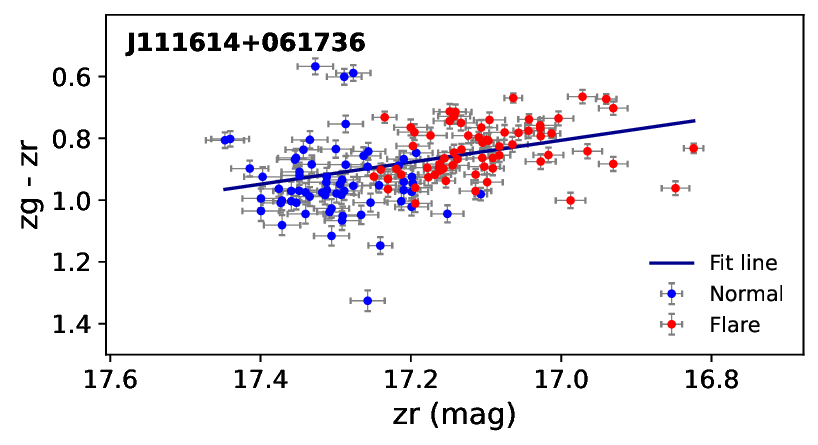}
        \includegraphics[width=0.49\textwidth]{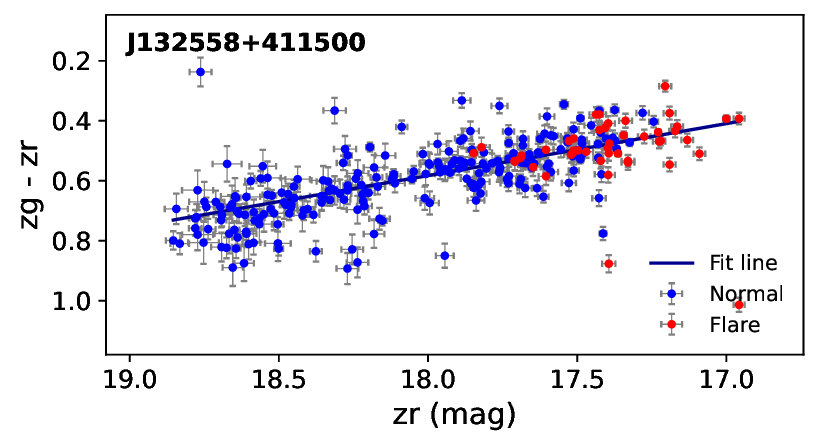}
\caption{CM diagrams of five not confirmed CL-AGN candidates: J075846+270515, J105344+492955, J111137+140118, J111614+061736, and J132558+411500.}
	\label{fig:nclcm}
\end{figure*}

\end{appendix}

\end{document}